\documentclass[twocolumn]{revtex4}
\usepackage[colorlinks,citecolor=blue]{hyperref}
\usepackage{textcomp}
\usepackage{subfigure}
\usepackage{verbatim}
\usepackage{bm}
\usepackage{hyperref}
\usepackage{amsmath}
\usepackage{graphicx}
\usepackage{epstopdf}
\usepackage{amssymb}
\usepackage{extarrows}
\usepackage{color}
\usepackage{CJK}
\usepackage{cancel}
\usepackage{ulem}
\newcommand{\ba}{\begin{eqnarray}}
\newcommand{\ea}{\end{eqnarray}}
\newcommand{\be}{\begin{equation}}
\newcommand{\ee}{\end{equation}}
\newcommand{\gr}{\mathrm{GR}}

\newcommand{\m}{\mathrm{max}}

\newcommand{\oct}{\mathrm{oct}}

\newcommand{\au}{\mathrm{AU}}

\newcommand{\IN}{\mathrm{in}}
\newcommand{\OUT}{\mathrm{out}}

\newcommand{\lk}{\mathrm{LK}}

\newcommand{\eff}{\mathrm{eff}}

\newcommand{\tot}{\mathrm{tot}}

\def\e1{e_1^2}

\begin{document}
\title{Merging Compact Binaries Near a Rotating Supermassive Black Hole: \\Eccentricity Excitation due to Apsidal Precession Resonance}
\author{Bin Liu$^{1}$, Dong Lai$^{1}$}
\affiliation{$^{1}$ Cornell Center for Astrophysics and Planetary Science, Cornell University, Ithaca, NY 14853, USA}

\begin{abstract}
We study the dynamics of merging compact binaries near a
rotating supermassive black hole (SMBH) in a hierarchical triple
configuration. We include various general relativistic effects that
couple the inner orbit, the outer orbit and the spin of the SMBH.
During the orbital decay due to gravitational radiation, the inner
binary can encounter an ``apsidal precession resonance" and experience
eccentricity excitation. This resonance occurs when the apsidal
precession rate of the inner binary matches that of the outer binary,
with the precessions driven by both Newtonian interactions and various
post-Newtonian effects. The eccentricity excitation requires the outer
orbit to have a finite eccentricity, and is most effective for triples
with small mutual inclinations, in contrast to the well-studied
Lidov-Kozai effect. The resonance and the associated eccentricity growth
may occur while the binary emits gravitational waves in the low-frequency
band, and may be detectable by future space-based gravitational wave
detectors.
\end{abstract}

\maketitle

\section{Introduction}

The detections of gravitational waves from merging binary black holes (BHs)
\cite{Abbott 2018a,Abbott 2018b,Zackay 2019a,Venumadhav 2019,Zackay 2019b}
have motivated many recent studies on their formation channels.
These include the traditional isolated binary evolution
\citep[e.g.,][]{Lipunov 1997,Lipunov 2007,Podsiadlowski 2003,Belczynski 2010,Belczynski 2016,Dominik 2012,Dominik 2013,Dominik 2015} and
chemically homogeneous evolution \citep[e.g.,][]{Mandel 2016,Marchant 2016},
gas-assisted mergers \citep[e.g.,][]{Bartos 2017}, and various flavors of dynamical channels that involve either
strong gravitational scatterings in dense clusters
\citep[e.g.,][]{Zwart (2000),OLeary (2006),Miller (2009),Banerjee (2010),Downing (2010),Ziosi (2014),Rodriguez (2015),Samsing (2017),
Samsing (2018),Rodriguez (2018),Gondan (2018)} or more gentle ``tertiary-induced mergers''
\citep[e.g.,][]{Blaes 2002,Miller 2002,Wen 2003,Antonini 2012,Antonini (2017),Silsbee (2017),Liu-ApJL,Liu-ApJ,Xianyu 2018,Hoang 2018,Liu ApJ III}.
Many recent studies have shown that merging BH and neutron star (NS) binaries can be formed efficiently
(via Lidov-Kozai (LK) oscillations \cite{Lidov,Kozai,Smadar}) with the aid of a tertiary body that moves on an
inclined (outer) orbit relative to the orbit of the inner binary.
Furthermore, the efficiency of the merger can be
enhanced when the triple is part of a quadruple system
\citep[e.g.,][]{Fang 2018,Liu Quadruple,Fragione Quadruple,Zevin 2019}, or more
generally, when the outer orbit experiences quasi-periodic external forcing
\citep[e.g.,][]{Hamers Dong Quad,Petrovich 2017,Fragione nulear cluster}.

In the standard LK-induced merger scenario, the leading order of post-Newtonian (PN) effect of general relativity (GR)
gives rise to apsidal precession of the inner binary, and this tends to
suppress LK oscillations or limit the maximum eccentricity \citep[e.g.,][]{Fabrycky 2007,Liu et al 2015,Anderson et al 2017}.
However, several numerical studies \citep[e.g.,][]{Ford,Mardling,Smadar PN}
based on secular triple equations (see \cite{C. M. Will PRD,C. M. Will PRL,Rodriguez (2020)}
for recent, more systematic study of the secular triple equations in PN theory)
also found evidence that with small mutual inclinations (no LK oscillations are allowed to occur),
significant eccentricity excitation might still be achieved under some conditions
(e.g., the clearest example of this phenomenon is displayed in Figure 14 of \cite{Ford}; \cite{Smadar PN}
added some other (generally non-essential) PN terms and called this ``GR-induced eccentricity excitation").
The physical explanation of the eccentricity growth at low inclinations in terms of ``apsidal procession resonance" was provided in \cite{Liu Yuan}
in the context of merging compact binaries
with tertiary (low-mass) companions. It was shown that a secular resonance occurs when the total apsidal precession
of the inner binary (driven by GR and the outer binary) matches the precession rate of the outer binary
(driven by the inner binary), and this resonance allows efficient ``transfer" of
eccentricity from the outer binary to the inner binary.

In this paper, we are interested in stellar-mass BH binary (BHB) mergers induced by a supermassive BH (SMBH).
Such BHBs may exist in abundance in the nuclear cluster (with a central SMBH)
due to various dynamical processes, i.e., scatterings and mass segregation \cite{OLeary 2009,Leigh 2018}.
Importantly, our recent study \cite{Liu nulear cluster} shows that several GR effects (including some of the ``cross terms" studied in
\cite{C. M. Will PRD}) introduced by a rotating SMBH
can generate extra precessions on the BH orbits, significantly increasing the merger fraction (as well as the merger rate).
This opens the question of how these GR effects modify the eccentricity growth mechanism due to the ``apsidal precession resonance".
We address this issue systematically in this paper.
We focus our attention to isolated BHB-SMBH triple systems, and do not consider other processes related to
scatterings and relaxation with surrounding stars in the cluster \cite{VanLandingham 2016,Petrovich 2017,Hamers 2018},
which may also change the character of SMBH-induced mergers of stellar BHBs.

Our paper is organized as follows.
In Section \ref{sec 2}, we review the essential GR effects (including the ``cross terms") relevant to BHB-SMBH triples
and present the secular equations in PN theory.
In Section \ref{sec 2 3}, we present some numerical examples to illustrate how secular apsidal resonance influences the orbital decay of BHB in triples.
In Section \ref{sec 3}, we perform analytical calculations for the
eccentricity excitation for coplanar systems residing near the resonance, and
explore the parameter space which can lead to the eccentricity excitation.
In Section \ref{sec 4}, we extend our calculations to systems with
slightly inclined outer binary and spin orientations.
We summarize our main results in Section \ref{sec 5}.

\section{Evolution of BHB near a SMBH}\label{sec 2}

\subsection{Equations for Standard LK-Induced Merger}\label{sec 2 1}

We consider a hierarchical triple system, composed of
an inner BH binary of masses $m_1$, $m_2$
and a distant companion (the SMBH) of mass $m_3$
that moves around the center of mass of the inner bodies.
The reduced mass for the inner binary is $\mu_\IN\equiv m_1m_2/m_{12}$, with $m_{12}\equiv m_1+m_2$.
Similarly, the outer binary has $\mu_\OUT\equiv(m_{12}m_3)/m_{123}$ with $m_{123}\equiv m_{12}+m_3$.
The semi-major axes and eccentricities are denoted by $a_\IN$, $a_\OUT$ and $e_\IN$, $e_\OUT$, respectively.
Therefore, the orbital angular momenta of two orbits are given by
$\textbf{L}_\IN=\mathrm{L}_\IN\hat{\textbf{L}}_\IN=\mu_\IN\sqrt{G m_{12}a_\IN(1-e_\IN^2)}\,\hat{\textbf{L}}_\IN$
and $\textbf{L}_\OUT=\mathrm{L}_\OUT\hat{\textbf{L}}_\OUT=\mu_\OUT\sqrt{G m_{123}a_\OUT(1-e_\OUT^2)}\,\hat{\textbf{L}}_\OUT$.
We define the mutual inclination between $\textbf{L}_\IN$ and $\textbf{L}_\OUT$ (inner and outer orbits) as $I$.
Throughout the paper, for convenience of notation, we will frequently omit the subscript ``$\IN$"
for the inner orbit.

To study the evolution of the merging inner BH binary under the influence of the tertiary companion,
we use the double-averaged (DA; averaging over both the inner and outer orbital periods)
secular equations of motion.
For the nearly coplanar systems studied in his paper,
the inner binary never reaches eccentricity close to unity, the DA approximation is valid
(see section 2.1 of \cite{Liu ApJ III}).
For the inner binary, the dynamics of the angular momentum $\textbf{L}$ and eccentricity $\mathbf{e}$ vectors are given by
\begin{eqnarray}
&&\frac{d \textbf{L}}{dt}=\frac{d \textbf{L}}{dt}\bigg|_{\mathrm{LK}}+\frac{d \textbf{L}}{dt}\bigg|_{\mathrm{GW}}~,\label{eq:Full Kozai 1}\\
&&\frac{d \mathbf{e}}{dt}=\frac{d \mathbf{e}}{dt}\bigg|_{\mathrm{LK}}+\frac{d \mathbf{e}}{dt}\bigg|_{\mathrm{GR}}+\frac{d
 \mathbf{e}}{dt}\bigg|_{\mathrm{GW}}~,\label{eq:Full Kozai 2}
\end{eqnarray}
where we include the contributions from the Newtonian potential of the external companion
(the first terms in Equations (\ref{eq:Full Kozai 1})-(\ref{eq:Full Kozai 2});
following the notation of \cite{Liu ApJ III}, we label these with the subscript ``LK"
since they can generate LK oscillations for sufficiently inclined orbits -- although in this paper we focus on non-LK regime),
the leading order PN correction,
and the dissipation due to gravitational waves (GW) emission.

The explicit DA equations of $d \textbf{L}/dt|_\mathrm{LK}$ and $d \textbf{e}/dt|_\mathrm{LK}$, are provided in \cite{Liu et al 2015}.
The Newtonian (LK) terms induce precession of eccentricity vectors on the timescale
\be
T_\lk=\frac{1}{\omega_\lk}=\frac{1}{n}\frac{m_{12}}{m_3}\bigg(\frac{a_{\OUT,\eff}}{a}\bigg)^3,
\ee
where $n=(G m_{12}/a^3)^{1/2}$ is the mean motion of the inner binary,
and $a_{\OUT,\eff}\equiv a_\OUT\sqrt{1-e^2_\OUT}$ is the effective outer binary separation.
Again, we label this $T_\lk$,
because for sufficiently inclined orbits, this is the LK timescale for oscillations in eccentricity and orbital inclination.

GR (1-PN correction) introduces pericenter precession of the inner binary,
\be\label{eq:e GR}
\frac{d \mathbf{e}}{dt}\bigg|_{\mathrm{GR}}=\omega_\mathrm{GR,in}\hat{\textbf{L}}\times\mathbf{e},
\ee
with the precession rate
\be\label{eq:GR}
\omega_\mathrm{GR,in}=\frac{3G^{3/2}m_{12}^{3/2}}{c^2a^{5/2}(1-e^2)}.
\ee
Similar equations apply to the outer binary, with
\ba
&&\frac{d \mathbf{e}_\OUT}{dt}\bigg|_{\mathrm{GR}}=\omega_\mathrm{GR,out}\hat{\textbf{L}}_\OUT\times\mathbf{e}_\OUT,\\
&&\omega_\mathrm{GR,out}=\frac{3G^{3/2}m_{123}^{3/2}}{c^2a_\OUT^{5/2}(1-e_\OUT^2)}.
\ea
During the LK oscillations, the short-range effect captured in Equation (\ref{eq:e GR}) plays a crucial role in determining the maximum
eccentricity $e_\m$ of the inner binary \citep[e.g.,][]{Fabrycky 2007}, that can be evaluated analytically
\citep[e.g.,][]{Liu et al 2015,Anderson et al 2017}.

Gravitational radiation draws energy and angular momentum from the BH orbit.
The rates of change of $\textbf{L}$ and $\mathbf{e}$ are given by \citep[]{Peters 1964}
\begin{eqnarray}
&&\frac{d \textbf{L}}{dt}\bigg|_{\mathrm{GW}}=-\frac{32}{5}\frac{G^{7/2}}{c^5}\frac{\mu^2 m_{12}^{5/2}}{a^{7/2}}
\frac{1+7e^2/8}{(1-e^2)^2}\hat{\textbf{L}},\label{eq:GW 1}\\
&&\frac{d \mathbf{e}}{dt}\bigg|_{\mathrm{GW}}=-\frac{304}{15}\frac{G^3}{c^5}\frac{\mu m_{12}^2}{a^4(1-e^2)^{5/2}}
\bigg(1+\frac{121}{304}e^2\bigg)\mathbf{e}.\label{eq:GW 2}
\end{eqnarray}
For reference, the merger time due to GW radiation of an isolated binary with the
initial semi-major axis $a_0$ and eccentricity $e_0$ is approximately given by
\ba\label{eq:Tmerger}
&&T_\mathrm{m}=T_\mathrm{m,0}(1-e_0^2)^{7/2}=\frac{5c^5 a_0^4}{256 G^3 m_{12}^2 \mu}(1-e_0^2)^{7/2}\\
&&\simeq10^{10}\bigg(\frac{60M_\odot}{m_{12}}\bigg)^2\bigg(\frac{15M_\odot}{\mu}\bigg)\bigg(\frac{a_0}{0.2\au}\bigg)^4(1-e_0^2)^{7/2}\mathrm{yrs}\nonumber.
\ea

Equations (\ref{eq:Full Kozai 1})-(\ref{eq:Full Kozai 2}), as well as the similar equations of motion of the outer binary (without GW emission),
completely determine the evolution of the triple system.
The \textit{Standard LK-Induced Merger }mechanism
(as studied in most previous works) considers sufficiently large mutual inclinations, and includes
the apsidal precession due to GR (Equation (\ref{eq:e GR})),
but neglects the GR effects associated with the rotating tertiary companion. This is adequate
when the tertiary mass $m_3$ is not much larger than the masses of the inner BHB.
However, for BHB-SMBH triples, with $m_3\gg m_1,m_2$, several
GR effects involving the SMBH can qualitatively change the efficiency and outcomes of LK-induced
mergers \cite{Liu nulear cluster}.

\subsection{Additional GR effects due to rotating SMBH}\label{sec 2 2}

For a rotating SMBH, the spin angular momentum is given by $\mathrm S_3=\chi_3G m_3^2/c$,
where $\chi_3\leqslant1$ is the Kerr parameter (we set $\chi_3=1$ in the numerical examples presented this paper).
In GR, the vectors $\textbf{L}$, $\textbf{L}_\OUT$, $\textbf{S}_3$, $\textbf{e}$ and $\textbf{e}_\OUT$ are coupled to each other,
inducing time evolution of these vectors. In a systematical post-Newtonian framework of triple dynamics
\cite{C. M. Will PRD,C. M. Will PRL,Rodriguez (2020)}, there are numerous terms.
We summarize the most essential effects below (also the leading-order effects).
The related equations are either from the classical work on binaries with spinning bodies \cite{Barker 1975}
(see also \cite{Racine 2008} and references therein), or can be derived (or extended to include eccentricity) ``by analogy",
i.e., by recognizing that the inner binary's orbital angular momentum
$\textbf{L}$ behaves like a ``spin".
As we see below, the vector forms of the equations we use are much more transparent than equations based on orbital elements
(see \cite{C. M. Will PRD,C. M. Will PRL,Rodriguez (2020)}), especially we are deal with misaligned $\textbf{L}$, $\textbf{L}_\OUT$ and $\textbf{S}_3$.

\textit{ Effect I: the coupling between $\textbf{L}_\OUT$ and $\textbf{S}_3$}.
In the BHB-SMBH system, the angular momentum of the outer binary $\textbf{L}_\OUT$
and the spin angular momentum $\textbf{S}_3$ of $m_3$
precesses around each other due to spin-orbit coupling if the two vectors
are misaligned (1.5 PN effect) \cite{Barker 1975,Fang Yun}:
\ba
\frac{d\textbf{L}_\OUT}{dt}\bigg|_\mathrm{L_\OUT S_3}=&&\omega_{\mathrm{L_\OUT S_3}}\hat{\textbf{S}}_3\times\textbf{L}_\OUT\label{eq:LOUT S3}, \\
\frac{d\textbf{e}_\OUT}{dt}\bigg|_\mathrm{L_\OUT S_3}=&&\omega_{\mathrm{L_\OUT S_3}}\hat{\textbf{S}}_3\times\textbf{e}_\OUT\nonumber\\
&&-3\omega_{\mathrm{L_\OUT S_3}}(\hat {\textbf{L}}_\OUT\cdot \hat {\textbf{S}}_3)\hat {\textbf{L}}_\OUT\times\textbf{e}_\OUT\label{eq:EOUT S3}, \\
\frac{d\hat{\textbf{S}}_3}{dt}\bigg|_\mathrm{S_3 L_\OUT}=&&\omega_{\mathrm{S_3 L_\OUT}}\hat{\textbf{L}}_\OUT\times\hat{\textbf{S}}_3,\label{eq:S3 S3}
\ea
where the orbit-averaged precession rates are
\be\label{eq:LOUT S3 rate}
\omega_\mathrm{L_\OUT S_3}=\frac{GS_3(4+3m_{12}/m_3)}{2c^2a_\OUT^3(1-e_\OUT^2)^{3/2}}=
\omega_{\mathrm{S_3 L_\OUT}}\frac{S_3}{L_\OUT}.
\ee
Since in our case, $\mathrm S_3$ can be easily larger than $\mathrm L_\OUT$, the de-Sitter precession (Equation (\ref{eq:S3 S3}))
is negligible.

\begin{figure*}
\centering
\begin{tabular}{cccc}
\includegraphics[width=9cm]{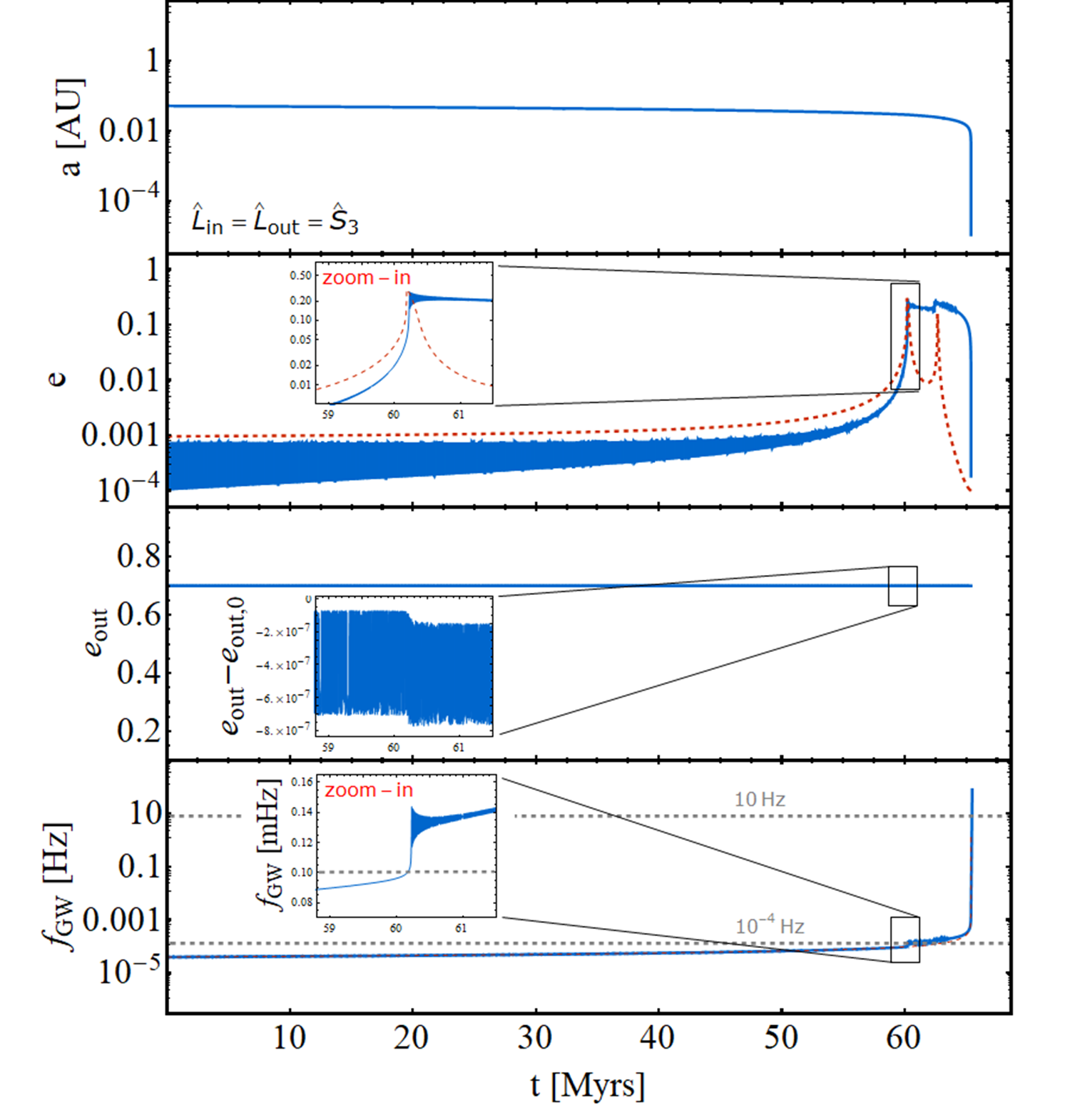}
\includegraphics[width=9cm]{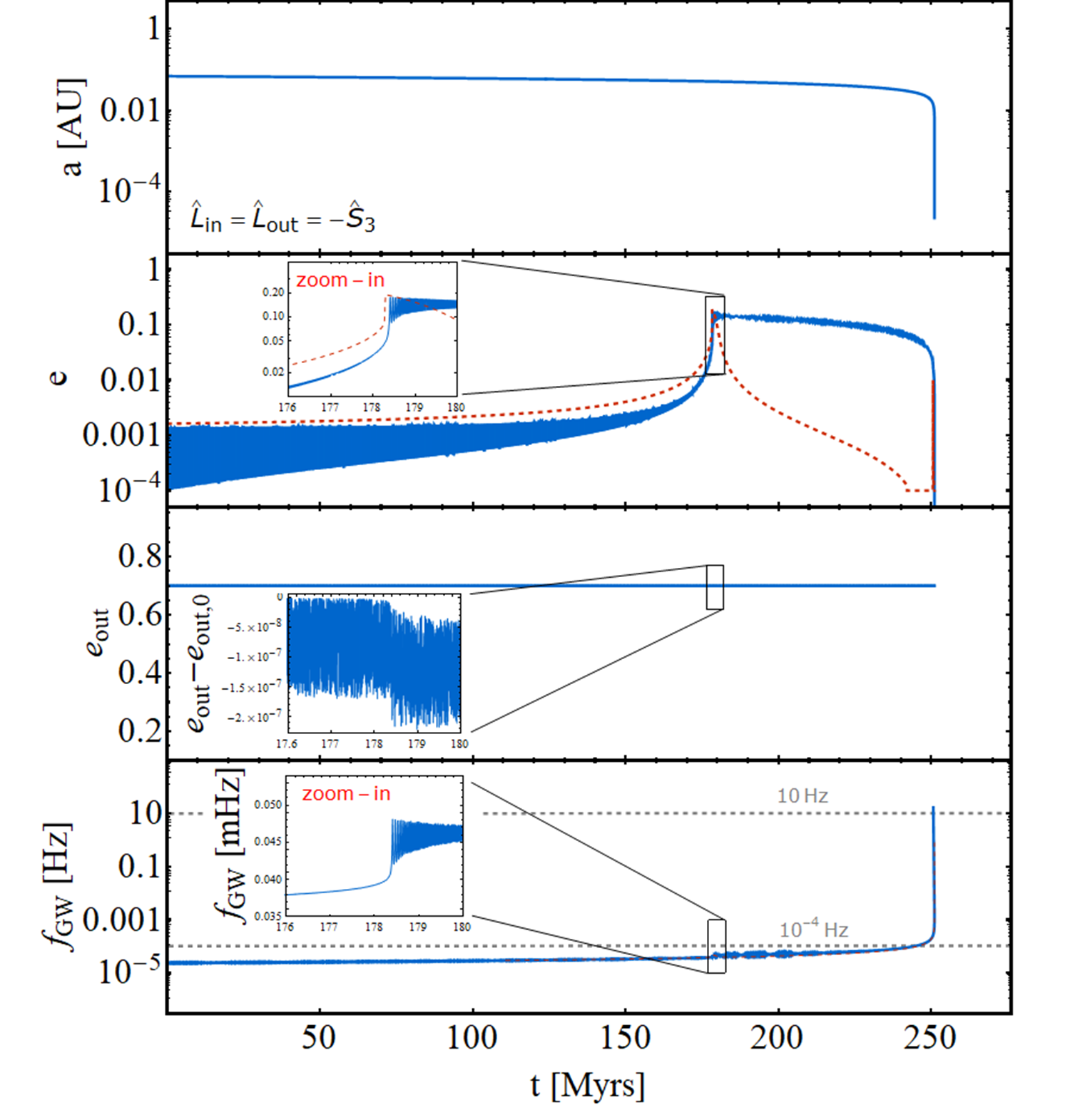}
\end{tabular}
\caption{The evolution of an inner merging BH-BH binary near a rotating SMBH
in a coplanar orbital configuration, with the SMBH spin either parallel (left) or anti-parallel (right) to the orbits..
The masses of BHs are $m_1=30M_\odot$, $m_2=20M_\odot$, and $m_3=10^8 M_\odot$, respectively.
The initial semimajor axes of the inner and outer binaries are $a_0=0.05\au$ (left panel), $a_0=0.07\au$ (right panel) and
$a_\OUT=90\au$, respectively. The eccentricities and longitudes of the
periapsis of two orbits are initialized as $e_0=10^{-4}$, $e_{\OUT,0}=0.7$ and $\varpi_0=\varpi_{\OUT,0}=0$, respectively.
The blue lines are obtained by the numerical integrations of the time evolution equations.
The red-dashed lines on the second panels depict the analytical maximum eccentricity of the inner binary
at the corresponding semi-major axis (see Section \ref{sec 3 1}).
The subfigures in the third and bottom panels show the difference of outer eccentricity ($e_\OUT-e_{\OUT,0}$) and
zoom-in of the GW peak frequency near the apsidal precession resonance.}
\label{fig:Orbital Evolution}
\end{figure*}

\textit{Effect II: the coupling between $\textbf{L}$ and $\textbf{L}_\OUT$}.
In addition to the Newtonian precession (driven by the tidal potential of $m_3$),
$\textbf{L}$ experiences an additional de-Sitter like (geodesic) precession
in the gravitational field of $m_3$ introduced by GR.
This is a 1.5 PN spin-orbit coupling effect, with $\textbf{L}$ behaving like a ``spin". We have
\ba
&&\frac{d \textbf{L}}{dt}\bigg|_\mathrm{L_\IN L_\OUT}=\omega_\mathrm{L_\IN L_\OUT}^{(\gr)}\hat{\textbf{L}}_\OUT\times\textbf{L},\label{eq:LinLout L}\\
&&\frac{d \textbf{e}}{dt}\bigg|_\mathrm{L_\IN L_\OUT}=\omega_\mathrm{L_\IN L_\OUT}^{(\gr)}\hat{\textbf{L}}_\OUT\times\textbf{e},\label{eq:LinLout e}
\ea
and the feedback from $\hat{\mathbf{L}}$, $\mathbf{e}$ on $\mathbf{L}_\OUT$ and $\mathbf{e}_\OUT$ are given by (see \cite{Barker 1975})
\ba
\frac{d \textbf{L}_\OUT}{dt}\bigg|_\mathrm{L_\OUT L_\IN}=&&\omega_\mathrm{L_\OUT L_\IN}^{(\gr)}
\hat{\textbf{L}}\times\textbf{L}_\OUT,\label{eq:LinLout Lout}\\
\frac{d \textbf{e}_\OUT}{dt}\bigg|_\mathrm{L_\OUT L_\IN}=&&\omega_\mathrm{L_\OUT L_\IN}^{(\gr)}\hat{\textbf{L}}\times\textbf{e}_\OUT
\label{eq:LinLout eout}\\
&&-3\omega_\mathrm{L_\OUT L_\IN}^{(\gr)}(\hat {\textbf{L}}_\OUT\cdot \hat {\textbf{L}})\hat {\textbf{L}}_\OUT\times\textbf{e}_\OUT, \nonumber
\ea
with
\be\label{eq:GR LinLout rate}
\omega_\mathrm{L_\IN L_\OUT}^{(\gr)}=\frac{3}{2}\frac{G (m_3+\mu_\OUT/3)n_\OUT}{c^2a_\OUT(1-e_\OUT^2)}
=\omega_\mathrm{L_\OUT L_\IN}^{(\gr)}\frac{L_\OUT}{L},
\ee
where $n_\OUT=(Gm_\tot/a_\OUT^3)^{1/2}$. Note the similarity between Equations (\ref{eq:LinLout Lout})-(\ref{eq:LinLout eout}) and
Equations (\ref{eq:LOUT S3})-(\ref{eq:EOUT S3}).

Note that both Equations (\ref{eq:LinLout L})-(\ref{eq:LinLout e}) are
required to keep $\textbf{L}\cdot\textbf{e}=0$.
Equations (\ref{eq:LinLout L})-(\ref{eq:LinLout e}) can also be reproduced through the ``cross terms"
in the PN equations of motion of hierarchical triple systems \cite{C. M. Will PRD,C. M. Will PRL}.

\textit{ Effects III: the coupling between $\textbf{L}$ and $\textbf{S}_3$}.
Since the semimajor axis of the inner orbit ($a$) is much smaller than the outer orbit ($a_\OUT$),
the inner binary can be treated as a single body approximately.
Therefore, the angular momentum $\hat {\textbf{L}}$ is coupled to the spin angular momentum $\textbf{S}_3$ of $m_3$,
and experiences Lens-Thirring precession. This is a 2 PN spin-spin coupling effect, with $\textbf{L}$ behaving like a ``spin". We have
\ba
\frac{d\textbf{L}}{dt}\bigg|_\mathrm{L_\IN S_3}=&&\omega_{\mathrm{L_\IN S_3}}\hat{\textbf{S}}_3\times\textbf{L}\nonumber\\
&&-3\omega_{\mathrm{L_\IN S_3}}(\hat {\textbf{L}}_\OUT\cdot \hat {\textbf{S}}_3)\hat {\textbf{L}}_\OUT\times\textbf{L}\label{eq:LIN S3},\\
\frac{d\textbf{e}}{dt}\bigg|_\mathrm{L_\IN S_3}=&&\omega_{\mathrm{L_\IN S_3}}\hat{\textbf{S}}_3\times\textbf{e}\nonumber\\
&&-3\omega_{\mathrm{L_\IN S_3}}(\hat {\textbf{L}}_\OUT\cdot \hat {\textbf{S}}_3)\hat {\textbf{L}}_\OUT\times\textbf{e}\label{eq:EIN S3}.
\ea
Note that Equation (\ref{eq:EIN S3}) is required to keep $\textbf{L}\cdot\textbf{e}=0$.
The back-reaction on the outer binary is gives (see Equations 64,65,70 of \cite{Barker 1975})
\ba
\frac{d\textbf{L}_\OUT}{dt}\bigg|_\mathrm{S_3 L_\IN}&&=-3\omega_{\mathrm{S_3 L_\IN}}
\Big[(\hat {\textbf{L}}_\OUT\cdot\hat {\textbf{L}})\hat {\textbf{S}}_3
+(\hat {\textbf{L}}_\OUT\cdot\hat {\textbf{S}}_3)\hat {\textbf{L}}\Big]\nonumber\\
&&\times\textbf{L}_\OUT,\label{eq:LOUT LIN S3}\\
\frac{d\textbf{e}_\OUT}{dt}\bigg|_\mathrm{S_3 L_\IN}&&=-3\omega_{\mathrm{S_3 L_\IN}}
\bigg\{(\hat {\textbf{L}}_\OUT\cdot\hat {\textbf{L}})\hat {\textbf{S}}_3+(\hat {\textbf{L}}_\OUT\cdot\hat {\textbf{S}}_3)\hat {\textbf{L}}\nonumber\\
&&+\Big[(\hat {\textbf{L}}\cdot\hat {\textbf{S}}_3)-5(\hat {\textbf{L}}_\OUT\cdot\hat {\textbf{L}})
(\hat {\textbf{L}}_\OUT\cdot\hat {\textbf{S}}_3)\Big]\hat {\textbf{L}}_\OUT\bigg\}\nonumber\\
&&\times\textbf{e}_\OUT.\label{eq:EOUT LIN S3}
\ea
In the above, the orbit-averaged precession rates are
\be\label{eq:LIN S3 rate}
\omega_\mathrm{L_\IN S_3}=\frac{GS_3}{2c^2a_\OUT^3(1-e_\OUT^2)^{3/2}}=\omega_{\mathrm{S_3 L_\IN}}\frac{L_\OUT}{L}.
\ee

Note $\omega_\mathrm{L_\IN S_3}/\omega_\mathrm{L_\IN L_\OUT}^{(\gr)}\sim (V_\OUT/c)\chi_3$
(where $V_\OUT$ is the orbital velocity of the outer binary and $\chi_3$
is the dimensionless spin parameter of the SMBH). Thus Effect III is important only when the outer binary is relativistic.

\section{Merging BHB with a coplanar SMBH: Numerical Examples}
\label{sec 2 3}

We begin with coplanar systems ($\hat{\mathbf{L}}=\hat{\mathbf{L}}_\OUT$)
with the SMBH spin either aligned ($\hat{\mathbf{S}}_3=\hat{\mathbf{L}}$) or anti-aligned
($\hat{\mathbf{S}}_3=-\hat{\mathbf{L}}$) with respect to the orbit.
Figure \ref{fig:Orbital Evolution} shows two examples.
All Newtonian (up to the octupole order) and GR effects discussed in Section \ref{sec 2} are included in the calculation.
In the left panels (with $\hat{\mathbf{L}}=\hat{\mathbf{L}}_\OUT=\hat{\mathbf{S}}_3$), the
eccentricity of the inner binary is negligible initially, but undergoes small-amplitude oscillations at the early phase, due to the
Newtonian perturbation from the outer binary. During the orbital decay, the eccentricity gets excited twice and
achieves a value of $\sim0.27$. This is the result of the ``apsidal precession resonance",
which allows the inner binary to efficiently ``gain" some eccentricity from the outer binary
(the eccentricity of the outer binary is initially $0.7$, and decreases only slightly as the system passes through the resonance)
(see more details in Section \ref{sec 3 2}).
The bottom panel shows the time evolution of the peak frequency of GW, given by \citep[][]{Wen 2003}
\be\label{eq:f peak}
f_\mathrm{GW} =\frac{(1+e)^{1.1954}}{\pi}\sqrt{\frac{G(m_{12})}{a^3(1-e^2)^3}}.
\ee
We see that the peak frequency is above $10^{-4}$ Hz at resonance,
and thus the system might be detected by the
future gravitational wave detectors operating at low frequencies.
After the resonances, the gravitational radiation reduces $e$ gradually,
circularizing the inner binary before the final merger.

For reference, the right panel of Figure \ref{fig:Orbital Evolution} shows the evolution for a system with
$\hat{\mathbf{L}}=\hat{\mathbf{L}}_\OUT=-\hat{\mathbf{S}}_3$.
A similar resonance feature occurs, although at different location (semi-major axis), and the inner binary eccentricity builds up to as high as $0.18$.

\section{Apsidal Precession Resonance: Analytical Results for Coplanar Systems}\label{sec 3}

For coplanar ($\hat{\mathbf{L}}=\hat{\mathbf{L}}_\OUT$), non-dissipative (with no gravitational radiation) systems,
the secular dynamics can be understood analytically. We first consider the case of small eccentricities, before
studying the finite eccentricity case.

\subsection{Linear (low-$e$) Systems}\label{sec 3 0}

For systems with $e$, $e_\OUT\ll1$, the evolution of $\mathbf{e}$ and $\mathbf{e}_\OUT$ is governed
by the linear Laplace-Lagrange equations \cite{MD}, with proper inclusion of the related GR precession terms \cite{Liu Yuan}.
Define the complex eccentricity variables
\be
\mathcal{E}_\IN=e_\IN \mathrm{exp}(i\varpi_\IN),
~~~\mathcal{E}_\OUT=e_\OUT \mathrm{exp}(i\varpi_\OUT),
\ee
where $\varpi_\IN$, $\varpi_\OUT$ are the longitude of pericenter of the inner and outer orbits.
The evolution equations are
\begin{align}\label{eq:eigen Equation}
\frac{d}{dt} \begin{pmatrix} \mathcal{E}_\IN \\ \mathcal{E}_\OUT  \end{pmatrix} &= i \begin{pmatrix} \omega_\IN & \nu_{\IN,\OUT}
\\ \nu_{\OUT,\IN} & \omega_\OUT \end{pmatrix}
\begin{pmatrix} \mathcal{E}_\IN \\ \mathcal{E}_\OUT \end{pmatrix},
\end{align}
where
\begin{eqnarray}
&&\omega_\IN=\frac{3}{4}n\frac{m_3}{m_{12}}\bigg(\frac{a}{a_\OUT}\bigg)^3\label{eq:A11}\\
&&~~~~~~+\omega_{\gr,\IN}+\omega_\mathrm{L_\IN L_\OUT}^{(\gr)}\mp2\omega_\mathrm{L_\IN S_3},\nonumber\\
&&\omega_\OUT=\frac{3}{4}n_\OUT\frac{m_1m_2}{m_{12}^2}\bigg(\frac{a}{a_\OUT}\bigg)^2\label{eq:A22}\\
&&~~~~~~~~+\omega_{\gr,\OUT}\mp2\omega_{\mathrm{L_\OUT S_3}}-2\omega_\mathrm{L_\OUT L_\IN}^{(\gr)}\pm6\omega_\mathrm{S_3 L_\IN}, \nonumber\\
&&\nu_{\IN,\OUT}=-\frac{15}{16}n\bigg(\frac{a}{a_\OUT}\bigg)^4\frac{m_3(m_1-m_2)}{m_{12}^2},\label{eq:A12}\\
&&\nu_{\OUT,\IN}=-\frac{15}{16}n_\OUT\bigg(\frac{a}{a_\OUT}\bigg)^3\frac{m_1m_2(m_1-m_2)}{m_{12}^3}.\label{eq:A21}
\end{eqnarray}
where in Equations (\ref{eq:A11})-(\ref{eq:A22}) the upper sign denotes the case of $\hat{\textbf{L}}=\hat{\textbf{L}}_\OUT=\hat{\textbf{S}}_3$,
and the lower sign the $\hat{\textbf{L}}=\hat{\textbf{L}}_\OUT=-\hat{\textbf{S}}_3$ case.

Starting with $e_0=0$, $e_\OUT=e_{\OUT,0}$ at $t=0$, Equation (\ref{eq:eigen Equation}) can be solved to determine the
time evolution of $e(t)$ \citep[see][]{Liu Yuan}. We find that $e(t)$ oscillates between
$e_0$ and $e_\m$, given by
\be\label{eq:e max}
e_\m=2e_{\OUT,0}\frac{|\nu_{\IN,\OUT}|}{\sqrt{(\omega_\IN-\omega_\OUT)^2+4\nu_{\IN,\OUT}\nu_{\OUT,\IN}}}.
\ee
Clearly, $e_\m$ attains the peak value when $\omega_\IN=\omega_\OUT$, at which
\ba\label{eq:e peak}
e_\mathrm{peak}&&=e_{\OUT,0}\bigg|\frac{\nu_{\IN,\OUT}}{\nu_{\OUT,\IN}}\bigg|^{1/2}
=e_{\OUT,0}\bigg|\frac{L'_\OUT}{L'_\IN}\bigg|^{1/2}\nonumber\\
&&=e_{\OUT,0}\frac{m_{12}^{3/4}}{m_{123}^{1/4}}\frac{m_3^{1/2}}{(m_1m_2)^{1/2}}\bigg(\frac{a_\OUT}{a}\bigg)^{1/4},
\ea
where $L'_\IN=L_\IN(e=0)$ and $L'_\OUT=L_\OUT(e_\OUT=0)$. We call this ``apsidal precession resonance".

The linear theory is valid only for $e\ll 1$ and $e_\OUT\ll 1$. Equation (\ref{eq:e peak}) shows that when $L'_\OUT\gg L'_\IN$,
as in the cases studied in this paper ($m_3\gg m_1, m_2$), even a small $e_{\OUT,0}$
may lead to unphysically larger $e_\mathrm{peak}$. In practice, Equation (\ref{eq:e peak}) is useful
in the sense that whenever it predicts $e_\mathrm{peak}$ of order unity of larger,
we can expect the inner binary to attain significant eccentricity, but the precise value of $e$ can only be obtained using nonlinear calculations,
as we discuss next.

\subsection{Finite Eccentricities}\label{sec 3 1}

For finite eccentricities, Equation (\ref{eq:eigen Equation}) breaks down, but the eccentricity evolution of a coplanar
triple can still be understood using energy and angular momentum conservations, without the need of
numerical integrations of the equations of motion.

The total energy of the triple can be written as
$\Phi_\tot=\Phi_\mathrm{N}+\Phi_\mathrm{extra}$, where
\be\label{eq:potential N}
\Phi_\mathrm{N}=\frac{\mu\Phi_0}{8}\bigg[-2-3e^2-\frac{15}{8}\varepsilon_{\oct} e (4+3e^2)\cos \Delta\varpi\bigg]
\ee
is the Newtonian potential energy between the inner and outer orbits in the octupole order \cite{Ford,Smadar OCT,Liu Yuan},
and $\Phi_\mathrm{extra}$ is the effective energy associated to the GR effects.
In Equation (\ref{eq:potential N}), $\Delta\varpi\equiv\varpi_\IN-\varpi_\OUT$,
with $\varpi_\IN$, $\varpi_\OUT$ the longitude of pericenters,
and
\be
\Phi_0\equiv\frac{G m_3 a^2}{a_\OUT^3(1-e_\OUT^2)^{3/2}}, ~~~
\varepsilon_{\oct}\equiv\frac{m_1-m_2}{m_{12}}\frac{a}{a_\OUT}\frac{e_\OUT}{1-e_\OUT^2}.
\ee

Various GR effects introduce extra apsidal precession
\begin{equation}\label{eq:extrae1}
\left.\frac{d\textbf{e}}{dt}\right|_{\mathrm{extra}}=\dot\omega_\mathrm{extra}~\hat{\textbf{L}}\times\textbf{e}.
\end{equation}
In terms of the effective potential, $\dot\omega_\mathrm{extra}$ is given by (see \cite{Fabrycky 2007} and Appendix \ref{Appendix A})
\begin{equation}\label{eq:extraomega}
\dot\omega_{\mathrm{extra}}=-\frac{\sqrt{1-e^2}}{e }\frac{1}{\mu\sqrt{Gm_{12}a}}\frac{\partial\Phi_{\mathrm{extra}}}{\partial e}.
\end{equation}
The effective $\Phi_\mathrm{extra}$ can be obtained from
\be\label{eq:potential extra}
\Phi_\mathrm{extra}=-\int\dot\omega_\mathrm{extra}\frac{e \mu\sqrt{Gm_{12}a}}{\sqrt{1-e^2}}de.
\ee
Equation (\ref{eq:extraomega}) is the canonical relations between Delaunay variables.
Equations (\ref{eq:extrae1})-(\ref{eq:potential extra}) can also apply to the outer binary. Consequently,
the potential energy associated with the periastron advance in the inner and outer orbits are given by \cite{Fabrycky 2007}
\ba
&&\Phi_{\gr,\IN}=-\frac{3G^2m_1m_2m_{12}}{c^2 a^2\sqrt{1-e^2}}\label{eq:potential GR},\\
&&\Phi_{\gr,\OUT}=-\frac{3G^2m_{12}m_3m_{123}}{c^2 a_\OUT^2\sqrt{1-e_\OUT^2}}.
\ea
Similarly, the potentials associated with the GR Effects I-III (Equations (\ref{eq:EOUT S3}), (\ref{eq:LinLout e}) and (\ref{eq:EIN S3}))
due to the rotating SMBH are
\ba
&&\Phi_\mathrm{L_\OUT S_3}=\pm\frac{G^2m_{12}(3m_{12}+4m_3)S_3}{2c^2 n_\OUT a_\OUT^4(1-e_\OUT^2)},\label{eq:potential Effect I}\\
&&\Phi_\mathrm{L_\IN L_\OUT}=\frac{G^3m_1m_2m_3(4m_{12}+3m_3)\sqrt{1-e^2}}{2c^2 n n_\OUT a a_\OUT^4(1-e_\OUT^2)},\label{eq:potential Effect II}\\
&&\Phi_\mathrm{L_\IN S_3}=\mp\frac{G^2m_1m_2S_3\sqrt{1-e^2}}{c^2naa_\OUT^3(1-e_\OUT^2)^{3/2}}\label{eq:potential Effect III}.
\ea
In Equations (\ref{eq:potential Effect I}) and (\ref{eq:potential Effect III}), the upper (lower) sign
refers to the $\hat{\textbf{S}}_3=\hat{\textbf{L}}$ ($\hat{\textbf{S}}_3=-\hat{\textbf{L}}$) case.
The detailed derivation of the potentials is presented in Appendix \ref{Appendix A}.

In the absence of dissipation (no GW), the total energy $\Phi_\mathrm{tot}$
(the sum of Equation (\ref{eq:potential N}) and (\ref{eq:potential GR})-(\ref{eq:potential Effect III})) is conserved.
This energy conservation, together with angular momentum conservation, $L_\mathrm{tot}=L+L_\OUT=\mathrm{constant}$,
completely determine the secular evolution of the triple system.

\begin{figure}
\centering
\begin{tabular}{cccc}
\includegraphics[width=8cm]{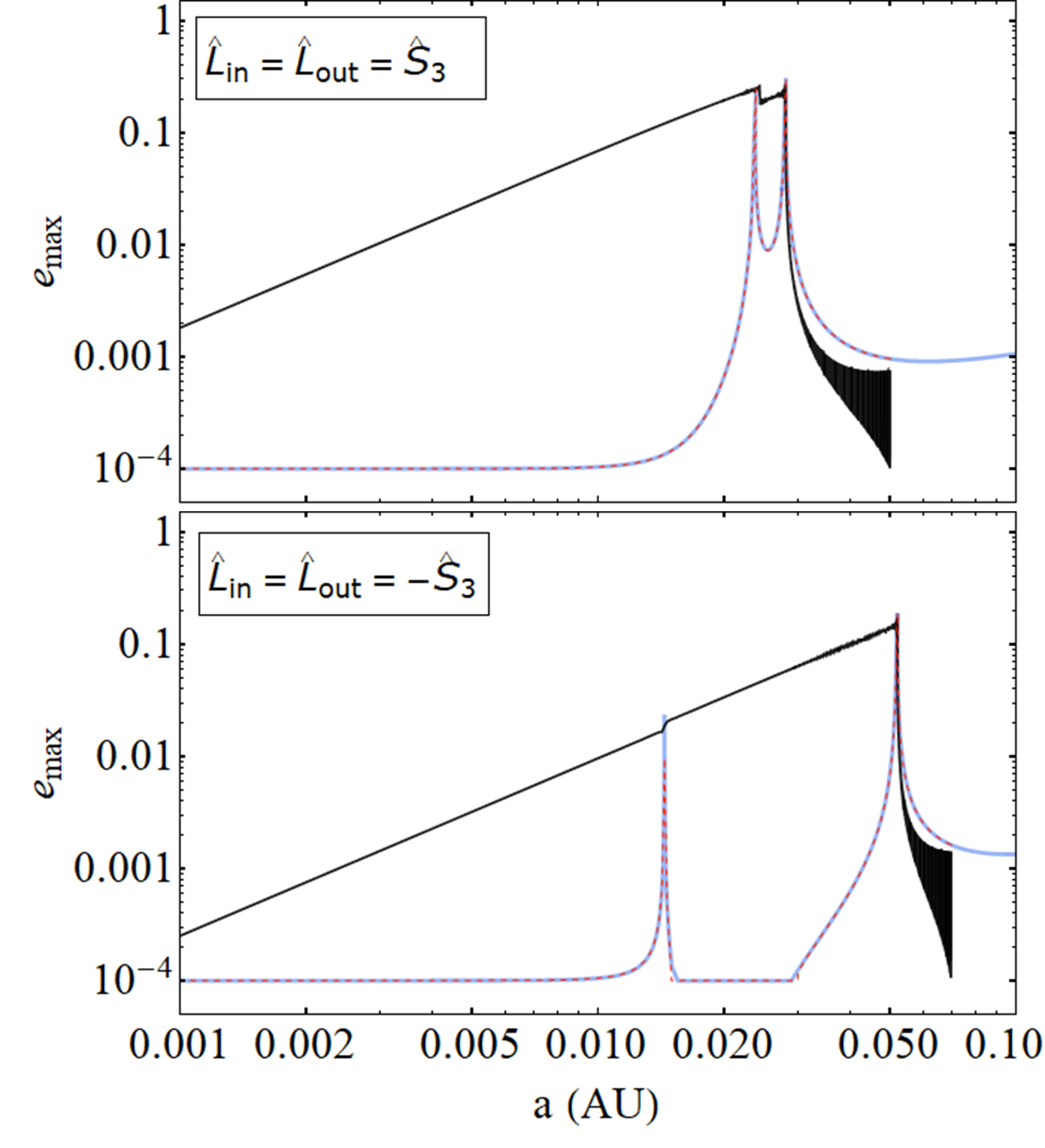}
\end{tabular}
\caption{The maximum eccentricity $e_\m$ of the inner binary as a function of the semimajor axis $a$ for systems starting with
$e_0=0$ and $e_{\OUT,0}=0.7$.
The system parameters are the same as in Figure \ref{fig:Orbital Evolution}.
The black and blue lines are the results from numerical integration with and without the GW emission.
The red dashed lines are achieved by the analytical formula according to the conservation laws. }
\label{fig:a e coplanar}
\end{figure}

\begin{figure}
\centering
\begin{tabular}{cccc}
\includegraphics[width=8cm]{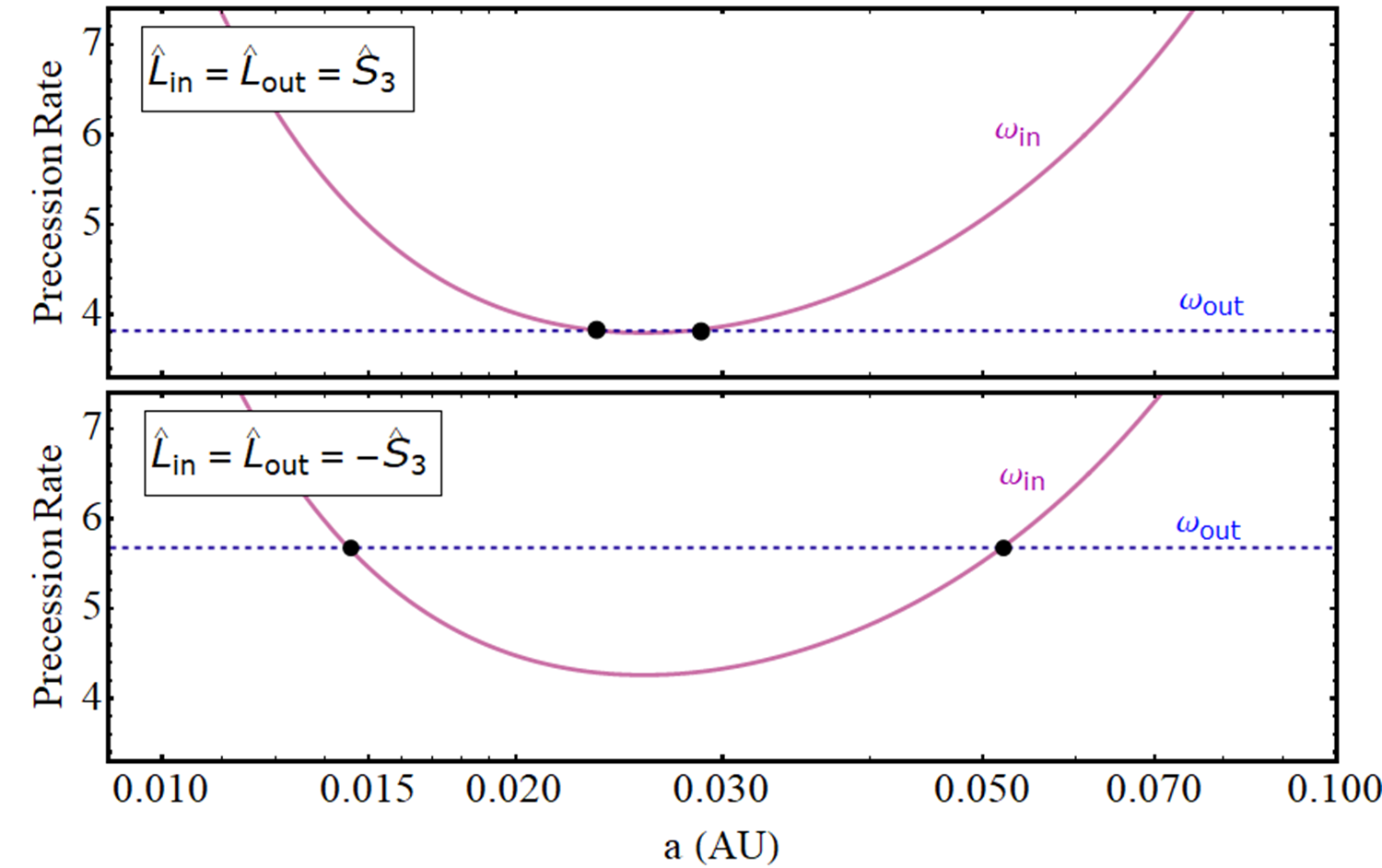}
\end{tabular}
\caption{The apsidal precession rates of the inner and outer binaries
(Equations \ref{eq:dot omega_IN} and \ref{eq:dot omega_OUT}) as a function of $a$.
The system parameters are the same as Figure \ref{fig:Orbital Evolution}.
The black dots represent the resonance location where $\omega_\IN=\omega_\OUT$.
}
\label{fig:resonance}
\end{figure}

\begin{figure*}
\centering
\begin{tabular}{cccc}
\includegraphics[width=17cm]{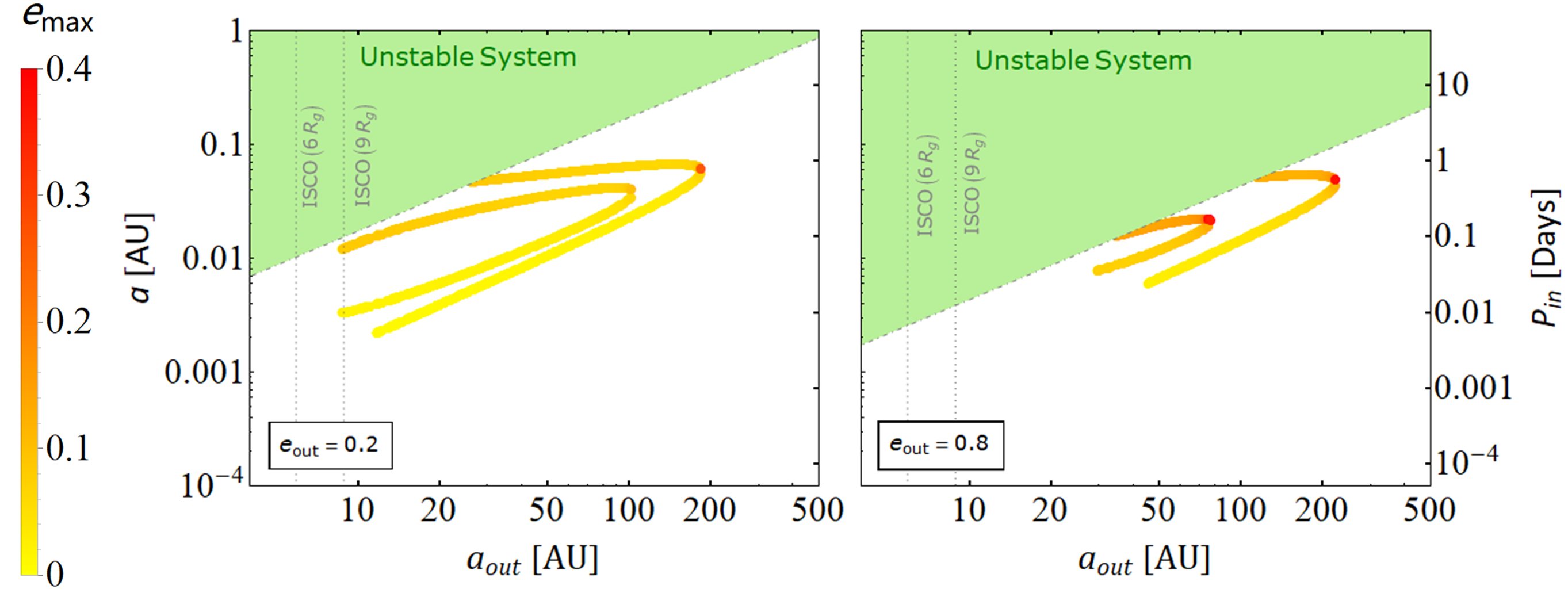}
\end{tabular}
\caption{Parameter space in the $a-a_\OUT$ plane, as well as the orbital period of the inner binary $P_\IN$-$a_\OUT$ plane,
where the apsidal precession resonance occurs.
The parameters are $m_1=30M_\odot$, $m_2=20M_\odot$, and $m_3=10^8M_\odot$.
The outer binary is set up with two different eccentricities, $e_\OUT=0.2$ for the left panel and $e_\OUT=0.8$ for the right panel.
The green region corresponds to the dynamically unstable triple systems.
The dotted line indicates the innermost stable circular orbits (ISCO) for the outer binary, where $R_g=(Gm_3)/c^2$.
The color-coded dots represent the values of the maximum eccentricity
of the inner binary (with negligible initial eccentricity)
due to the resonance, as calculated by
energy and angular momentum conservations.
Here, the inner and outer curves are from the triple systems with aligned and anti-aligned $\hat{\textbf{S}}_3$, respectively.}
\label{fig:parameter space a aout}
\end{figure*}

Suppose $e=0$ and $e_\OUT=e_{\OUT,0}$ at $t=0$, we can use the conservation of $L_\mathrm{tot}$ and $\Phi_\mathrm{tot}$
to determine $e_\m$, the maximum eccentricity attained by the inner binary. Note that since $L_\OUT\gg L$,
conservation of $L_\mathrm{tot}$ implies
\be\label{eq:eout Taylor}
e_\OUT\simeq e_{\OUT,0}+\frac{1-e_{\OUT,0}^2}{e_{\OUT,0}}\frac{L-L_0}{L_{\OUT,0}},
\ee
where $L_0$ and $L_{\OUT,0}$ are the initial values of $L$ and $L_\OUT$, respectively.
Solving the energy and angular momentum conservation laws
(Equation (\ref{eq:eout Taylor})) yields $e$ as a function of $\Delta\varpi$.
The maximum eccentricity $e_\m$ is achieved at either $\Delta \varpi=0$ or $\pi$,
depending on the initial value of $\Delta\varpi$, and whether $\Delta\varpi$ librates or circulates.

Figure \ref{fig:a e coplanar} shows $e_\m$ obtained by different methods as a function of $a$
where the parameters are the same as in Figure \ref{fig:Orbital Evolution}.
Here, the black solid lines are from the numerical integrations including the GW emission, while the
the blue solid lines are achieved by integrating the equations without the GW radiation.
We evolve the system for about $\sim10^3$ yrs for a given semimajor axis $a$ and record $e_\m$ during the evolution.
The red dashed lines are the analytical prediction using the two conservation laws.

The numerical result (without the GW) are in a good agreement with the analytical calculation.
We see that there are two evident peaks of $e_\m$ in the blue and red dashed lines, indicating the resonance arises twice
(see also Figure \ref{fig:resonance}).
However, in the ``real" evolution of the system (i.e. with GW),
the eccentricity preserves the memory of the excitation: Once the growth in $e$ happens, the orbit can only be circularized by GW emission gradually.

\subsection{Eccentricity Excitation in coplanar System}\label{sec 3 2}

\begin{figure*}
\centering
\begin{tabular}{cccc}
\includegraphics[width=15.7cm]{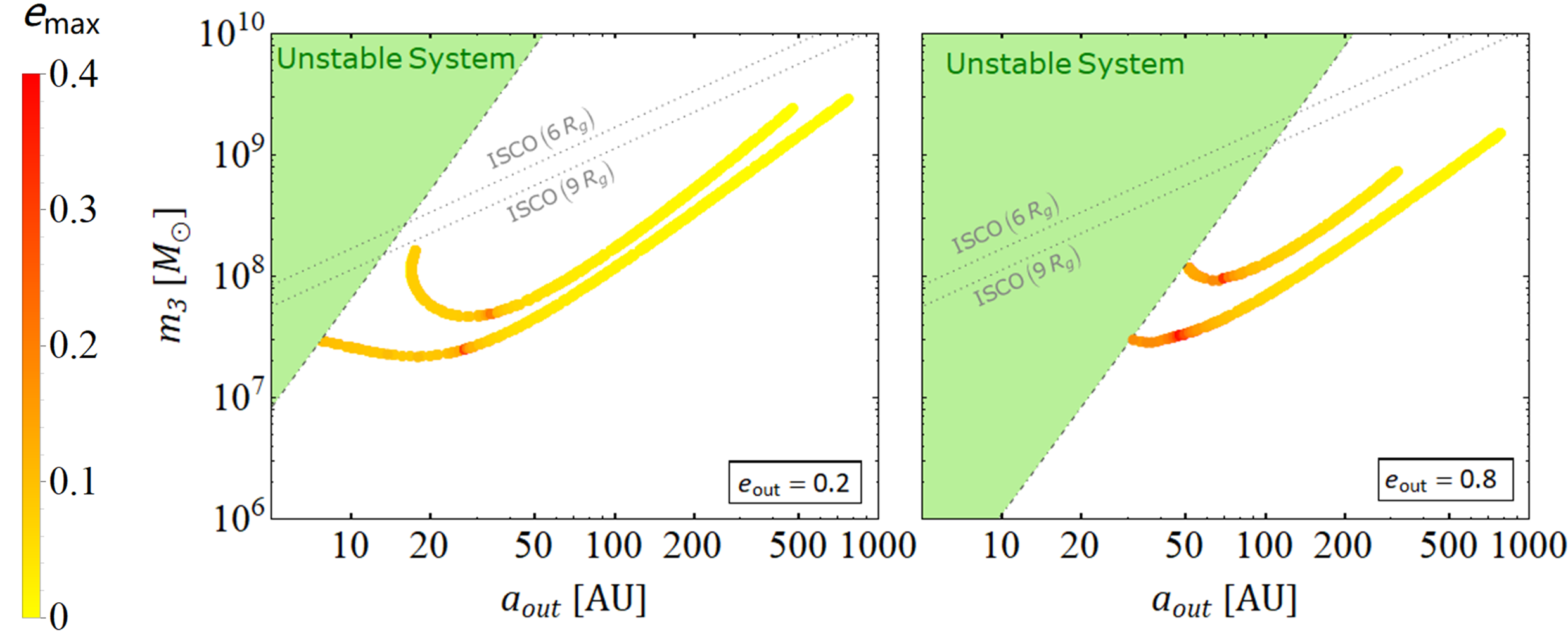}
\end{tabular}
\caption{Same as Figure \ref{fig:parameter space a aout}, but in the $m_3-a_\OUT$ plane, for $a=0.02$ AU.}
\label{fig:parameter space m3 aout}
\end{figure*}

Our analysis in Section \ref{sec 3 0} for linear (low-$e$) systems shows that the inner binary
attains a peak eccentricity at the resonance $\omega_\IN=\omega_\OUT$. For finite eccentricities, the resonance is
not precise, but we expect a similar peak eccentricity occurs when $\omega_\IN=\omega_\OUT$, with
\ba
&&\omega_\IN=\omega_{\lk,\IN}+\omega_{\gr,\IN}+\omega_\mathrm{L_\IN L_\OUT}^{(\gr)}\mp2\omega_\mathrm{L_\IN S_3}
\label{eq:dot omega_IN},\\
&&\omega_\OUT=\omega_{\gr,\OUT}\mp2\omega_{\mathrm{L_\OUT S_3}}-2\omega_\mathrm{L_\OUT L_\IN}^{(\gr)}\pm6\omega_\mathrm{S_3 L_\IN}
\label{eq:dot omega_OUT},
\ea
where the various frequencies include the dependence of finite $e_\OUT\simeq e_{\OUT,0}$.
Note that in Equation (\ref{eq:dot omega_OUT}), the last two terms are much smaller than the corresponding terms in
Equation (\ref{eq:dot omega_IN}) since $L_\IN/L_\OUT \ll 1$.

Figure \ref{fig:resonance} shows $\omega_\IN$ and $\omega_\OUT$ as a function of $a$ for the examples depicted in Figure \ref{fig:Orbital Evolution}.
When $a$ is relatively large, the Newtonian interaction is strong and $\omega_{\lk,\IN}$
is the dominated source in $\omega_\IN$. So we have $\omega_\IN\gg\omega_\OUT$.
On the other hand, near the merger of the inner binary ($a$ is small), the GR effect becomes important and
dominate the precession, leading to $\omega_\IN\ll\omega_\OUT$. In between, we see that
$\omega_\IN$ crosses $\omega_\OUT$ twice, creating two ``apsidal precession resonances" during the
orbital decay.

\subsection{Parameter Space}\label{sec 3 3}

\begin{figure*}
\centering
\begin{tabular}{cccc}
\includegraphics[width=13cm]{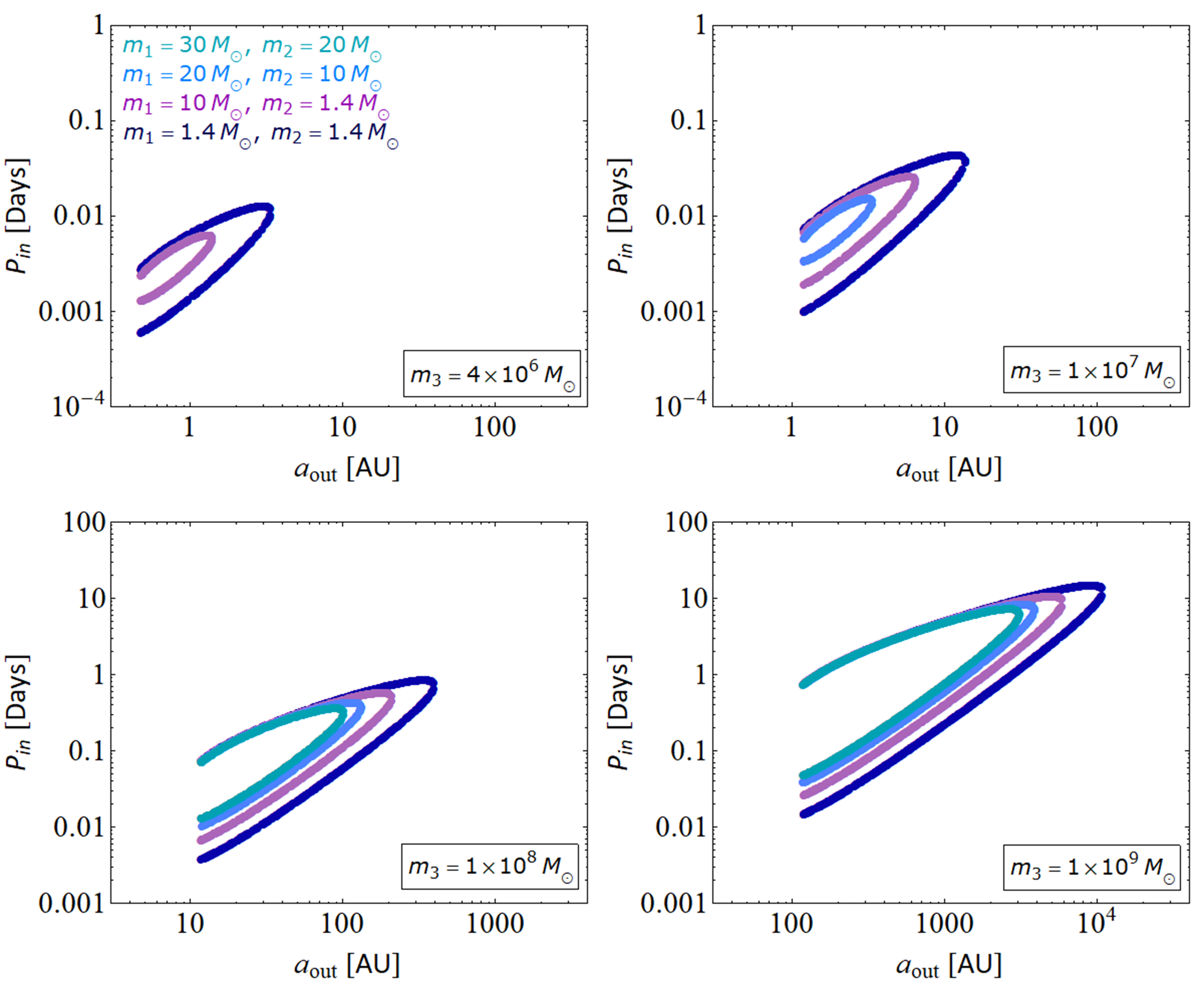}
\end{tabular}
\caption{Similar to Figure \ref{fig:parameter space a aout}, but for the coplanar triple systems with aligned $\hat{\textbf{S}}_3$.
Here, we only show the systems satisfying the stability criterion and outside the ISCO.
The outer eccentricity $e_\OUT$ is set to be $0.5$ in all cases. We select different BH binary masses and SMBH mass, as labeled.}
\label{fig:parameter space a aout BHNS}
\end{figure*}

\begin{figure}
\centering
\begin{tabular}{cccc}
\includegraphics[width=8cm]{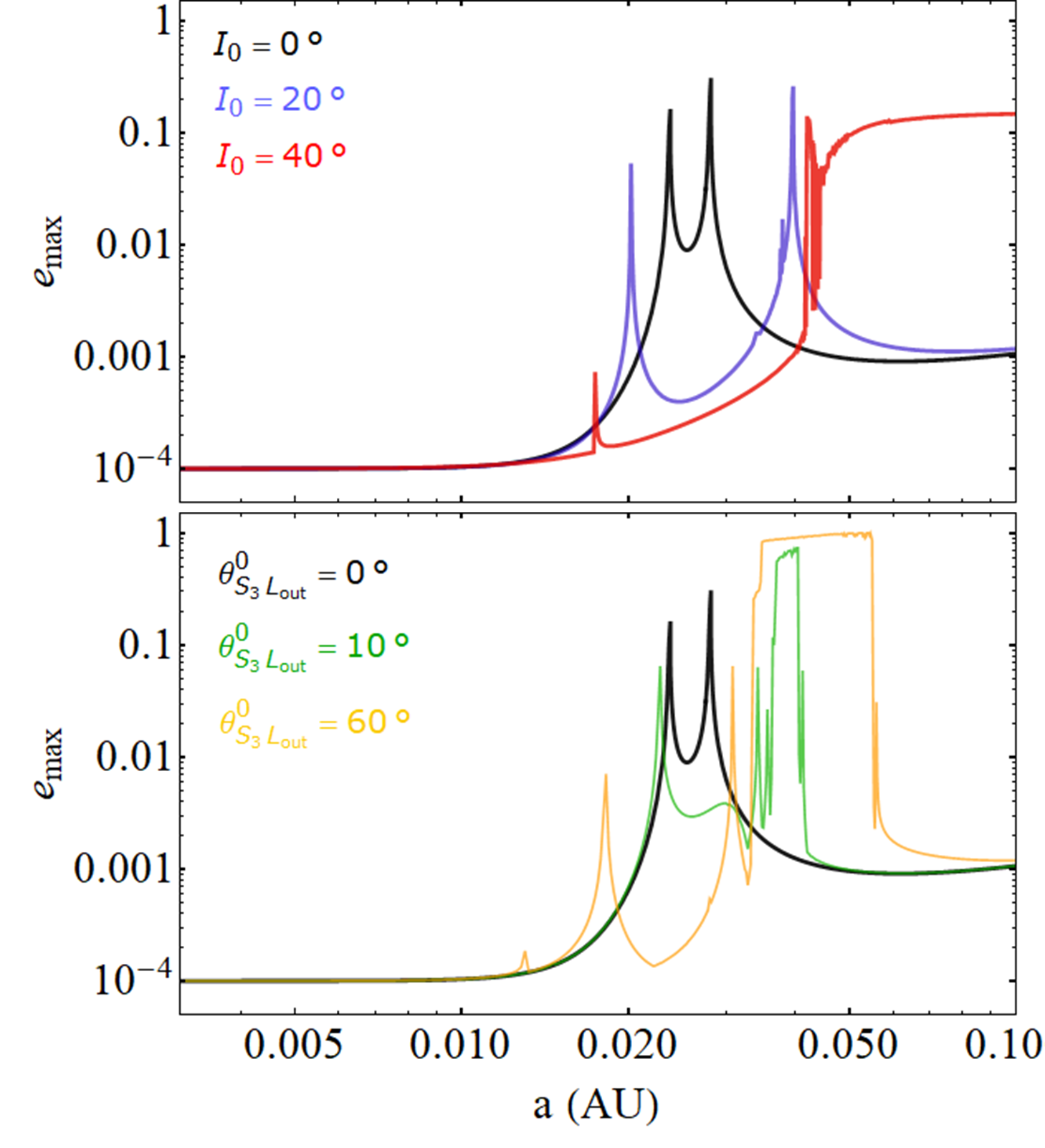}
\end{tabular}
\caption{The maximum eccentricity $e_\m$ as a function of the inner binary semimajor axis $a$, obtained by integrating
the octupole equations of motion and including the GR effects (but no GW emission). The upper panel shows the results
from various initial inclinations (as labeled) and aligned $\hat{\textbf{S}}_3$ (parallel to $\hat{\textbf{L}}_\OUT$). The lower panel shows the
results from different initial $\hat{\textbf{S}}_3-\hat{\textbf{L}}_\OUT$ misalignment angles (as labeled) and coplanar tripla.
The other parameters are the same as in Figure \ref{fig:Orbital Evolution}.}
\label{fig:a e inclined}
\end{figure}

\begin{figure}
\centering
\begin{tabular}{cccc}
\includegraphics[width=9cm]{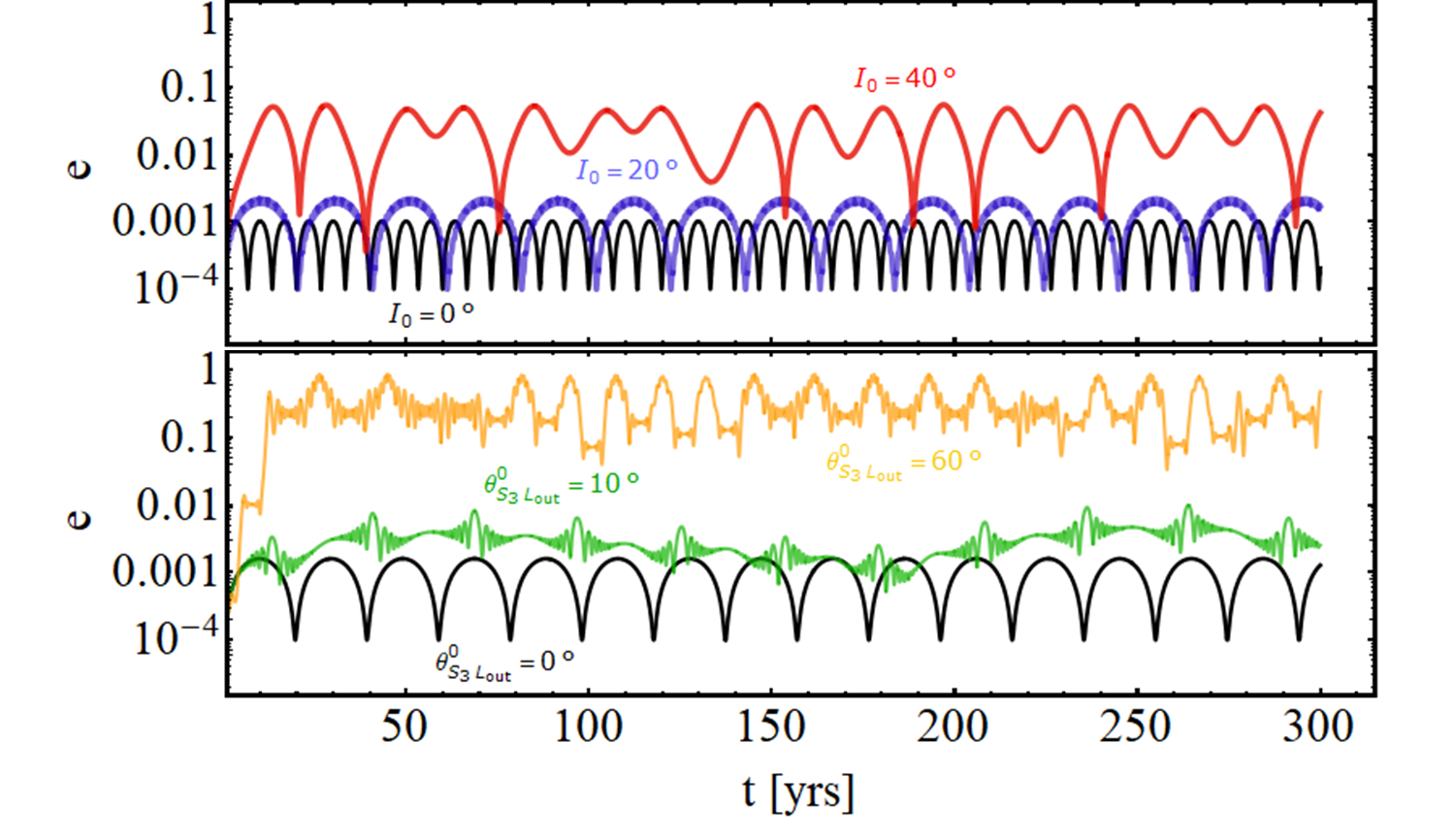}
\end{tabular}
\caption{Evolution examples of the orbital eccentricity in the inner binary.
The parameters are picked from the cases in Figure \ref{fig:a e inclined}, and $a=0.046\au$ (upper panel) and $a=0.0363\au$ (lower panel).}
\label{fig:Orbital Evolution no GW}
\end{figure}

For a given set of system parameters, the criterion of apsidal precession resonance ($\omega_\IN=\omega_\OUT$)
provides a good estimation on the resonance radius
(resonance occurs at the location $a=a_{\IN,\mathrm{Res}}$ for a given $a_\OUT$; or $a_\OUT=a_{\OUT,\mathrm{Res}}$ for a given $a$).
Combining the analytical analysis in Section \ref{sec 3 1}, we can predict the maximum eccentricity of the inner binary at resonance.

Figure \ref{fig:parameter space a aout} illustrates the level of eccentricity excitation
in the $a (P_\IN)-a_\OUT$ plane for the given $m_3$ and $e_\OUT$.
The green region corresponds to the space where the triple system is dynamically unstable, where
the dot-dashed line is the instability limit according to \cite{Kiseleva}.
The dotted lines indicate the innermost stable circular orbits (ISCO) for the outer binary,
where $R_g=(G m_3)/c^2$ (the ISCO ranges from $R_g$ to $9R_g$ depending on the spin magnitude and orientation relative to the orbit).
The color-coded dots denote the locations ($a_{\IN,\mathrm{Res}}$) at resonance and the values of $e_\m$.
The inner part apply to the systems with $\hat{\textbf{L}}=\hat{\textbf{L}}_\OUT=\hat{\textbf{S}}_3$
and outer part $\hat{\textbf{L}}=\hat{\textbf{L}}_\OUT=-\hat{\textbf{S}}_3$.

\begin{figure*}
\centering
\begin{tabular}{cccc}
\includegraphics[width=9cm]{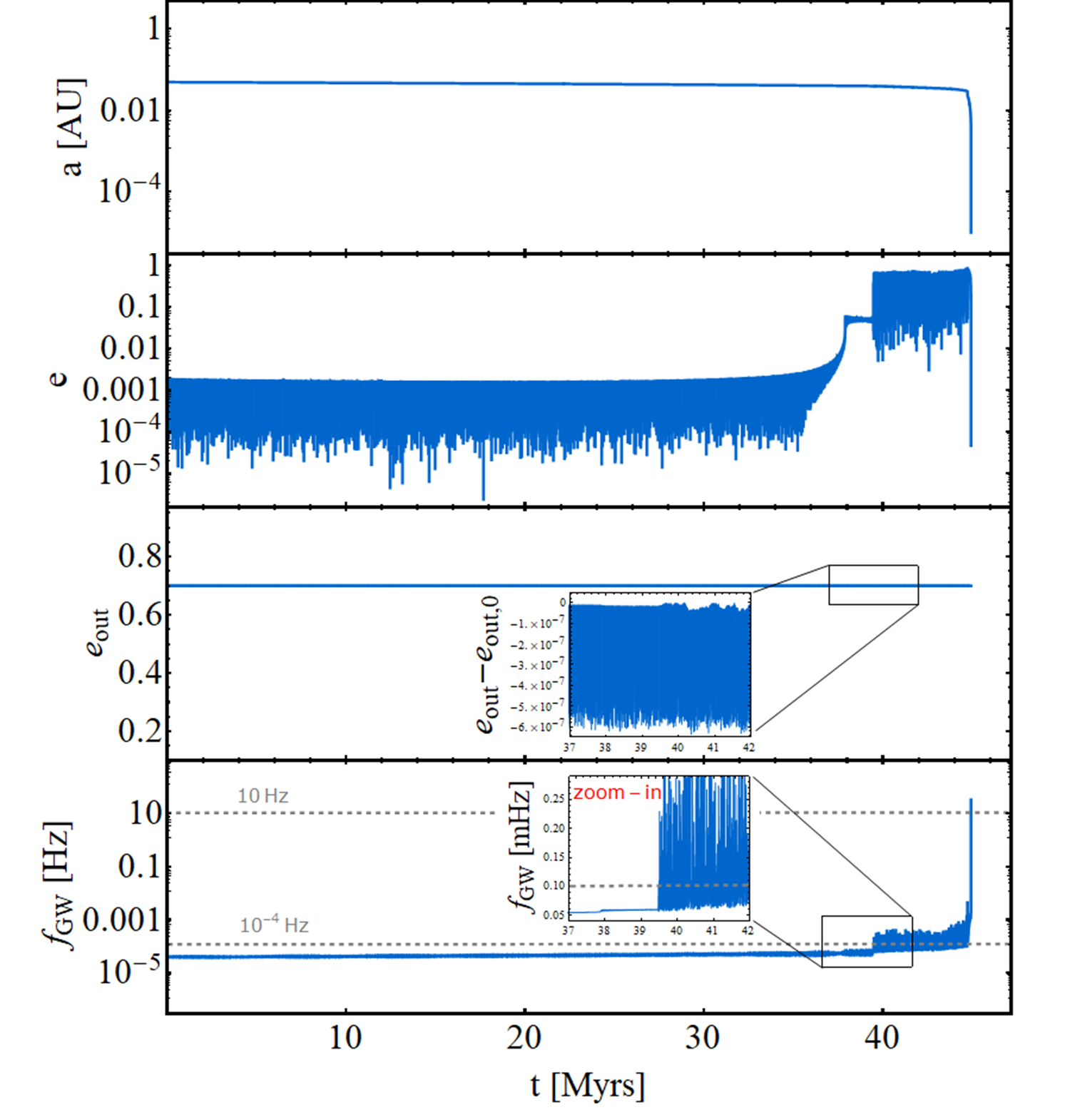}
\includegraphics[width=9cm]{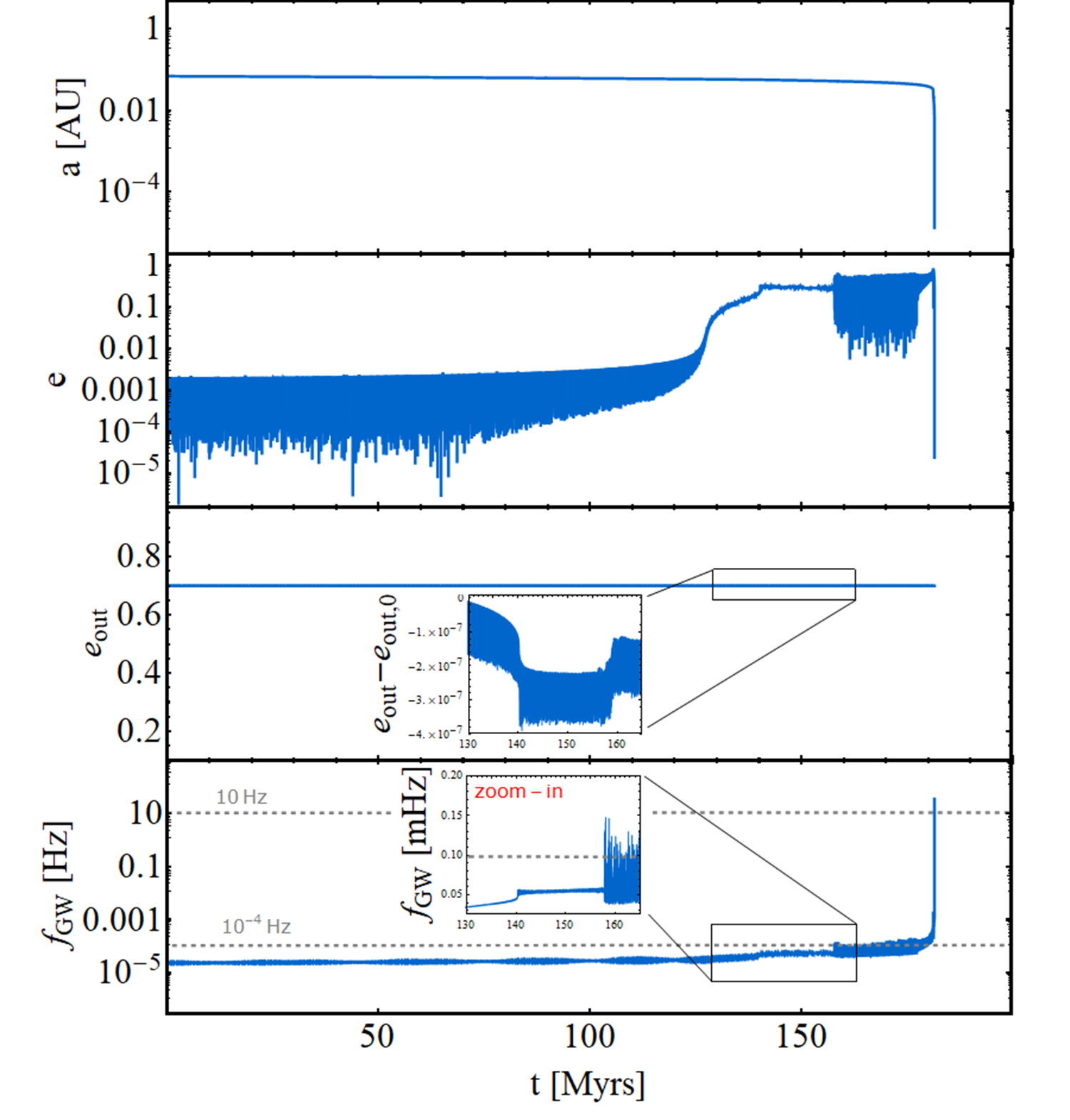}
\end{tabular}
\caption{Similar to Figure \ref{fig:Orbital Evolution}, but for triple systems with finite initial binary inclinations
and $\hat{\textbf{S}}_3-\hat{\textbf{L}}_\OUT$ misalignment angles: $I_0=10^\circ$, $\theta^0_{\mathrm{S_3L_\OUT }}=10^\circ$ (left)
and $\theta^0_{\mathrm{S_3L_\OUT }}=170^\circ$ (right).
The other parameters are the same as in Figure \ref{fig:Orbital Evolution}. The qualitative behavior of the inclined systems is similar
to the coplanar systems, but the maximum eccentricities due to resonance are larger.}
\label{fig:Orbital Evolution inclined}
\end{figure*}

We see that in the left panel of Figure \ref{fig:parameter space a aout}, for the resonance to occur, the BHB must be close to the SMBH.
In the case of aligned $\hat{\textbf{S}}_3$, two locations of $a_{\IN,\mathrm{Res}}$ tend to converge as $a_\OUT$ increases.
The maximum $e_\m$ is achieved at the largest $a_\OUT$.
Furthermore, the eccentricity of the outer binary is crucial to produce significant $e_\m$
(as shown in the right panel).

Figure \ref{fig:parameter space m3 aout} shows the similar results but in the $m_3-a_\OUT$ plane (with $a$ fixed at $0.02$ AU).
The upper and lower color-coded dots are from the case of aligned and anti-aligned $\hat{\textbf{S}}_3$, respectively.
We find that the apsidal precession resonance is expected for a wide range of $m_3$.
In particular, small $m_3$ leads to larger values of $e_\m$
for a given $a$ and $e_\OUT$.

Figure \ref{fig:parameter space a aout BHNS} displays the resonance locations for four types of inner binaries (as labeled) in $P_\IN-a_\OUT$ plane,
with four different SMBH masses.
Note that we only restrict to the configuration where $\hat{\textbf{S}}_3$ is aligned with $\hat{\textbf{L}}_\OUT$, and
have considered the criteria of stability \cite{Kiseleva} and ISCO.
We see that, in terms of the occurrence of resonances, the small mass SMBH (top left panel) favors the low mass binary,
especially the BH-NS binary and NS-NS binary.
To produce resonance, these binaries must be concentrated in the inner-most region ($\sim$AU) around the SMBH.
Since the BHB emits GW in the low frequency range before chirping to the LIGO band,
if the time evolution on the binary eccentricity could be measured (e.g. sources near the Galactic center),
it would provide a useful test for the binary formation channels.
As the mass of the SMBH increases, the apsidal precession resonance can emerge for the high mass binaries,
and the regions of interest are located farther from the SMBH,
as shown in other panels of Figure \ref{fig:parameter space a aout BHNS}.

\section{Resonance in the inclined Systems}\label{sec 4}

In this section, we extend our exploration to the general cases of mutually inclined inner/outer binaries, and misaligned spin of SMBH
with respect to $\hat{\textbf{L}}_\OUT$. Since no simple analytical result can be derived, we
sample $a$ to determine the resonance location numerically for given outer binary parameters.

We first consider the same example as in Figure \ref{fig:Orbital Evolution} (with aligned $\hat{\textbf{S}}_3$),
but increase the initial inclination to the higher values. We integrate the secular equations of motion
with no GW radiation, evolving the system for $\sim10^3$ yrs, and record $e_\m$ during
the evolution. The results are shown in the upper panel of Figure \ref{fig:a e inclined}.
Compared to the fiducial example (black lines; the same as in Figure \ref{fig:a e coplanar}),
we see that the resonance locations change and spread for larger inclinations.
As $I_0$ approaches $40^\circ$, the dynamics is largely determined by the LK oscillations
when $a\gtrsim0.05$ AU (For smaller $a$'s, LK oscillations are suppressed),
and the eccentricity excitation due to resonance tends to be erased.

We can see several examples of the eccentricity evolution in the upper panel of Figure \ref{fig:Orbital Evolution no GW},
where $a$ is fixed to be $0.046$ AU. The maximum eccentricity varies.
The irregular behavior in the case of $I_0=40^\circ$ arises due to the combined influences of Effects I-III.

Alternatively, we vary the spin orientation of the SMBH,
choosing the initial misalignment angle between $\hat{\textbf{S}}_3$ and $\hat{\textbf{L}}_\OUT$ (i.e., $\theta^0_{\mathrm{S_3L_\OUT }}$) to be
$0^\circ$, $10^\circ$ and $60^\circ$,  while keeping the mutual binary inclination angle to zero.
The results are shown in the lower panel of Figure \ref{fig:a e inclined}. Somewhat surprisingly,
in this case, $e_\m$ is sensitive to $\theta^0_{\mathrm{S_3L_\OUT }}$ and it can grow to larger values, approaching the unity.
This is because Effect I plays a crucial role, and
LK oscillations can be triggered (especially for $\theta^0_{\mathrm{S_3L_\OUT }}=60^\circ$)
due to an inclination resonance \citep[e.g.,][]{Liu nulear cluster}.
To illustrate the $e$ evolution,
in the lower panel of Figure \ref{fig:Orbital Evolution no GW}, we see the growth of $e$ becomes chaotic
for the fixed $a=0.0363$ AU.

We now include the dissipative effect of gravitational radiation. If we adopt the slightly inclined outer binary and
aligned (or anti-aligned) $\hat{\textbf{S}}_3$, the system can still encounter the apsidal precession resonance.
Figure \ref{fig:Orbital Evolution inclined} shows an example of a merging BHB with the inclined SMBH ($I_0=10^\circ$).
The initial misalignment angles between $\hat{\textbf{L}}_\OUT$ and $\hat{\textbf{S}}_3$ are
$10^\circ$ (left panel) and $170^\circ$ (right panel), respectively.
We find that the behaviors of $e$ and $e_\OUT$, in particular the excitation of the inner binary eccentricity, are
more significant than the case of coplanar triple. When the resonance is encountered during the orbital decay, $e$ increases while $e_\OUT$
decreases. The systems have the maximum eccentricities in excess of $0.81$.

If the triple systems are initialized with arbitrary $I_0$ and $\theta^0_{\mathrm{S_3L_\OUT }}$, the
evolution may become chaotic, and the eccentricity of the inner binary can easily to grow to close to unity \citep[e.g.,][]{Liu nulear cluster}.
Such ``GR-enhanced" channel may play an important role in BHB mergers.
A comprehensive parameter space study is beyond the scope of this paper and
we leave it to a future work.

\section{Discussion and Conclusion}\label{sec 5}

In this paper, we have studied the dynamics of compact BH-BH binaries
under the influence of a nearby rotating SMBH in a hierarchical triple
configuration. We have presented the general secular equations of
motion that govern the evolution of the (BH-BH)-SMBH triple system,
including various general relativistic (GR) effects that couple the
inner and outer orbits and the spin of the SMBH (Section \ref{sec 2}). These
post-Newtonian equations of motion are derived and extended from
previous work on binaries with spinning bodies \cite{Barker 1975}. In our
recent work \cite{Liu nulear cluster}, we have shown that several of
these GR effects can significantly influence the rate of tertiary
induced binary mergers via Lidov-Kozai mechanism. In this paper, we
focus on systems with small mutual inclinations such that Lidov-Kozai
oscillation does not happen. We show that compact binaries near a
SMBH can experience an ``apsidal precession resonance'', where the
pericenter precession rate of the inner binary matches that of the
outer binary. Both precessions are driven by the combined effects of
Newtonian gravitational interaction and general relativity. The
resonance results in an efficient ``transfer'' of eccentricity from
the outer binary to the inner binary, leading to eccentricity growth
of the inner binary. An example of the eccentricity growth due to
apsidal precession resonance during binary merger near a SMBH is shown
in in Figure \ref{fig:Orbital Evolution}.

We provide analytical analysis for coplanar systems with small
eccentricities (linear theory; Section \ref{sec 3 0}) and finite
eccentricities (non-linear theory; Section \ref{sec 3 1}). The linear
theory gives a useful criterion for the resonance (Equations
(\ref{eq:e max}), (\ref{eq:dot omega_IN}) and (\ref{eq:dot omega_OUT})), but the non-linear theory is needed to accurately
predict the value of maximum eccentricity excitation (see Figure \ref{fig:a e coplanar}). The growth of the eccentricity in the inner
binary at resonance can be understood as ``angular momentum exchange''
(i.e., eccentricity exchange) between the inner and outer
binaries. For the systems studied in this paper ($m_{\rm SMBH}=m_3\gg m_1, m_2$),
even a weakly eccentric outer orbit (SMBH's orbit) can
excite appreciable eccentricity in the inner BH-BH binary, and the
peak eccentricity increases as the outer binary becomes more eccentric
(see Figure \ref{fig:parameter space a aout}).

The eccentricity growth due to apsidal resonance generally operates
for triple systems with small mutual inclinations (in contrast to
Lidov-Kozai oscillations, which require high mutual inclinations) and
allows for generally misaligned spin orientations of the SMBH.
(see Figure \ref{fig:a e inclined} and \ref{fig:Orbital Evolution no GW}).
The GR effects (especially Effect I; see Equations (\ref{eq:LOUT S3})-(\ref{eq:LOUT S3 rate})) play an
important role, and can make the orbital evolution of the BH binary
chaotic. The eccentricity of the merging binary can attain a larger
value compared to the corresponding coplanar case (see Figure \ref{fig:Orbital Evolution inclined}).

The apsidal precession resonance can occur while the binary is emitting
gravitational waves in the low-frequency band, thus potentially
detectable by future gravitational wave detectors operating at low
frequencies, such as LISA, DECIGO \cite{DECIGO} and TianQin \cite{TianQin}.
Binary mergers near the Galactic Center would be of great interest, particularly if the
eccentricity evolution can be tracked.
Of course, whether stellar BBHs  actually exist so close to the SMBH is unknown.
Such close-in BBHs may result from mass segregation and the scatterings. A SMBH could also tidally capture a BHB to a bound orbit, and
the BHB would inspiral towards the SMBH in a close orbit due to gravitational radiation.
A complete study of this topic is beyond the scope of this paper.
On the other hand, the apsidal precession resonance may play a role in the
scenario of BH binary merger in the Active Galactic Nuclei disk
\citep[e.g.,][]{Bartos 2017}. In this case, a BH binary aligns its
orbit with the disk, and moves to the migration trap close to
the central SMBH.  The final configuration of (BH-BH)-SMBH system may then satisfy the criterion of
resonance studied in this paper.

\section{Acknowledgments}

This work is supported in part by the NSF grant AST-1715246 and NASA
grant NNX14AP31G.

\appendix
\section{The Effective Potentials of GR Effects}
\label{Appendix A}

In the hierarchical coplanar triple system, the secular Hamiltonian is given by
\ba
\mathcal{H}&&=\mathcal{H}_1+\mathcal{H}_2 + \Phi\nonumber\\
&&=-\frac{G m_1m_2}{2a}-\frac{G m_{12}m_3}{2a_\OUT}+\Phi_\mathrm{N}+\Phi_\mathrm{extra},
\ea
where $\Phi_\mathrm{N}$ is given by Equation (\ref{eq:potential N}), and $\Phi_\mathrm{extra}$ is the effective potential due to various GR effects.
For the coplanar systems ($\hat {\textbf{L}}=\hat {\textbf{L}}_\OUT$), we have $\mathcal{H}=\mathcal{H}(\varpi_\IN, L_\IN, \varpi_\OUT, L_\OUT)$, where
$\varpi_\IN$ and $\varpi_\OUT$ are the longitude of pericenter of the inner and outer binaries, respectively,
and $L_\IN=\mu\sqrt{G m_{12}a(1-e^2)}$, $L_\OUT=\mu_\OUT\sqrt{G m_{123}a_\OUT(1-e_\OUT^2)}$ are the respective canonical momenta.
The canonical equations of motion for the inner and outer orbits are
\ba
&&\dot\varpi_\IN=\frac{\partial \mathcal{H}}{\partial L_\IN}, ~~~\dot L_\IN=-\frac{\partial \mathcal{H}}{\partial \varpi_\IN}\label{eq:H in}, \\
&&\dot\varpi_\OUT=\frac{\partial \mathcal{H}}{\partial L_\OUT}, ~~~\dot L_\OUT=-\frac{\partial \mathcal{H}}{\partial \varpi_\OUT}\label{eq:H out}.
\ea
Since $\mathcal{H}$ depends on $\varpi_\IN$ and $\varpi_\OUT$ through
($\varpi_\IN-\varpi_\OUT$) (see Equation (\ref{eq:potential N})),
the second equations in (\ref{eq:H in}) and (\ref{eq:H out}) yield angular momentum conservation
\be
L_\IN + L_\OUT ={\rm constant}.
\ee
In vector form, the first equations in (\ref{eq:H in}) and (\ref{eq:H out}) imply that the apsidal precession due to $\Phi_\mathrm{extra}$ is given by
\ba
\frac{d\textbf{e}}{dt}\bigg|_\mathrm{extra}&&=\dot\varpi_{\mathrm{extra}}^{(\IN)}\hat {\textbf{L}}\times\textbf{e}\\
\frac{d\textbf{e}_\OUT}{dt}\bigg|_\mathrm{extra}&&=\dot\varpi_{\mathrm{extra}}^{(\OUT)}\hat {\textbf{L}}_\OUT\times\textbf{e}_\OUT,
\ea
where
\be
\dot\varpi_{\mathrm{extra}}^{(\IN)}=\frac{\partial\Phi_\mathrm{extra}}{\partial L_\IN},~~~
\dot\varpi_{\mathrm{extra}}^{(\OUT)}=\frac{\partial\Phi_\mathrm{extra}}{\partial L_\OUT}.
\ee
Thus, for a given apsidal precession rate, the ``common" extra potential can be evaluated by
\be\label{eq:extrapotantial common}
\Phi_\mathrm{extra}=\int\dot\varpi_{\mathrm{extra}}^{(\IN)}dL_\IN=
\int\dot\varpi_{\mathrm{extra}}^{(\OUT)}dL_\OUT.
\ee

~~Consider the various GR effects discussed in Section \ref{sec 2}.
For Effect I, Equation (\ref{eq:EOUT S3}) reduces to
\be\label{eq:EOUT S3 Coplanar}
\frac{d\textbf{e}_\OUT}{dt}\bigg|_\mathrm{L_\OUT S_3}=
(\mp~2\varpi_{\mathrm{L_\OUT S_3}})\hat {\textbf{L}}_\OUT\times\textbf{e}_\OUT.
\ee
The upper (lower) sign
refers to the $\hat{\textbf{S}}_3=\hat{\textbf{L}}$ ($\hat{\textbf{S}}_3=-\hat{\textbf{L}}$) case.
Using (\ref{eq:LOUT S3 rate}), the effective potential can be obtained:
\ba
\Phi_\mathrm{L_\OUT S_3}&&=\int(\mp~2\varpi_{\mathrm{L_\OUT S_3}})\mu_\OUT\sqrt{Gm_{123}a_\OUT}~d\sqrt{1-e_\OUT^2}\nonumber\\
&&=\pm\frac{G^2m_{12}(3m_{12}+4m_3)S_3}{2c^2 n_\OUT a_\OUT^4(1-e_\OUT^2)}.
\ea
For Effect II, Equations (\ref{eq:LinLout e}) and (\ref{eq:LinLout eout}) reduce to
\ba
\frac{d \textbf{e}}{dt}\bigg|_\mathrm{L_\IN L_\OUT}=&&\varpi_\mathrm{L_\IN L_\OUT}^{(\gr)}\hat{\textbf{L}}_\OUT\times\textbf{e},
\label{eq:LinLout e Appendix}\\
\frac{d \textbf{e}_\OUT}{dt}\bigg|_\mathrm{L_\OUT L_\IN}=&&-2\varpi_\mathrm{L_\OUT L_\IN}^{(\gr)}
\hat{\textbf{L}}_\OUT\times\textbf{e}_\OUT. \label{eq:LinLout eout Appendix}
\ea
Uding Equation (\ref{eq:GR LinLout rate}), the common effective potential can be obtained:
\ba
\Phi_\mathrm{L_\IN L_\OUT}&&=\int\varpi_\mathrm{L_\IN L_\OUT}^{(\gr)}\mu\sqrt{Gm_{12}a}~d\sqrt{1-e^2}\nonumber\\
&&=\int(-2\varpi_\mathrm{L_\OUT L_\IN}^{(\gr)})\mu_\OUT\sqrt{Gm_{123}a_\OUT}~d\sqrt{1-e_\OUT^2}\nonumber\\
&&=\frac{G^3m_1m_2m_3(4m_{12}+3m_3)\sqrt{1-e^2}}{2c^2 n n_\OUT a a_\OUT^4(1-e_\OUT^2)}.
\ea
For Effect III, Equations (\ref{eq:EIN S3}) and (\ref{eq:EOUT LIN S3}) reduce to
\ba
\frac{d\textbf{e}}{dt}\bigg|_\mathrm{L_\IN S_3}=&&(\mp~2\varpi_{\mathrm{L_\IN S_3}})\hat {\textbf{L}}_\OUT\times\textbf{e},
\label{eq:Ein S3 Appendix}\\
\frac{d\textbf{e}_\OUT}{dt}\bigg|_\mathrm{S_3 L_\IN}&&=(\pm~6\varpi_{\mathrm{S_3 L_\IN}})\hat {\textbf{L}}_\OUT\times\textbf{e}_\OUT.
\label{eq:Eout S3 Appendix}
\ea
Again, the upper (lower) sign
refers to the $\hat{\textbf{S}}_3=\hat{\textbf{L}}$ ($\hat{\textbf{S}}_3=-\hat{\textbf{L}}$) case.
Using (\ref{eq:LIN S3 rate}), the effective potential is given by
\ba
\Phi_\mathrm{L_\IN S_3}&&=\int(\mp~2\varpi_{\mathrm{L_\IN S_3}})\mu\sqrt{Gm_{12}a}~d\sqrt{1-e^2}\nonumber\\
&&=\int(\pm~6\varpi_{\mathrm{S_3 L_\IN}})\mu_\OUT\sqrt{Gm_{123}a_\OUT}~d\sqrt{1-e_\OUT^2}\nonumber\\
&&=\mp\frac{G^2m_1m_2S_3\sqrt{1-e^2}}{c^2naa_\OUT^3(1-e_\OUT^2)^{3/2}}.
\ea

\end{document}